\documentclass[aps,prb,twocolumn,groupedaddress,showpacs,floatfix]{revtex4}
\usepackage{graphics}
\usepackage{epsfig}
\begin{document}
\title{Diagrammatic perturbation theory and the pseudogap}

\author{P. Monthoux$^\dagger$}
\affiliation{Cavendish Laboratory, University of Cambridge
\\Madingley Road, Cambridge CB3 0HE, United Kingdom}

\begin{abstract}

We study a model of quasiparticles on a two-dimensional
square lattice coupled to Gaussian distributed dynamical
molecular fields. We consider two types of such fields,
a vector molecular field that couples to the quasiparticle 
spin-density and a scalar field coupled to the quasiparticle
number density. The model describes quasiparticles coupled 
to spin or charge fluctuations, and is solved by a Monte
Carlo sampling of the molecular field distributions.
The nonperturbative solution is compared to various
approximations based on diagrammatic perturbation theory.
When the molecular field correlations are sufficiently
weak, the diagrammatic calculations capture the qualitative 
aspects of the quasiparticle spectrum.
For a range of model parameters near the magnetic boundary, 
we find that the quasiparticle spectrum is qualitatively 
different from that of a Fermi liquid, in that it shows
a double peak structure, and that the diagrammatic 
approximations we consider fail to reproduce, 
even qualitatively, the nonperturbative
results of the Monte Carlo calculations. This suggests
that the magnetic pseudogap induced by a coupling to
antiferromagnetic spin-fluctuations and the spin-splitting
of the quasiparticle peak induced by a coupling to
ferromagnetic spin-fluctuations lie beyond
diagrammatic perturbation theory.
While a pseudogap opens when quasiparticles are coupled 
to antiferromagnetic fluctuations, such a pseudogap is 
not observed in the corresponding charge-fluctuation case 
for the range of parameters studied, where vertex corrections 
are found to effectively reduce the strength of the interaction.
This suggests that one has to be closer to the border
of long-range order to observe pseudogap effects in the
charge-fluctuation case than for a spin-fluctuation
induced interaction under otherwise similar conditions.
The diagrammatic approximations that contain first order
vertex corrections show the enhancement of the spin-fluctuation
induced interaction and the suppression of the effective
interaction in the charge-fluctuation case. However, for
the range of model parameters considered here, the multiple
spin or charge-fluctuation exchange processes not included
in the diagrammatic approximations considered are found to 
be important, especially for quasiparticles coupled to 
charge fluctuations.

\end{abstract}
\date{\today}
\maketitle

\section{Introduction}

The concept of elementary excitations and the diagrammatic
perturbation-theoretic methods borrowed from quantum field 
theory have given us, over the past decades, many powerful 
insights into the behavior of materials. In a number of 
cases, however, these concepts and methods don't seem to work. 
In previous papers\cite{PM1,PhilMag}, we presented results 
on a nonperturbative extension of the magnetic interaction 
model, which had until then been extensively used in the 
context of diagrammatic approaches. These latter applications 
were successful in many respects: in the Eliashberg approximation,
the magnetic interaction model correctly anticipated the pairing 
symmetry of the Cooper state in the copper oxide 
superconductors\cite{DwavePrediction} and is consistent 
with spin-triplet p-wave pairing in superfluid $^3He$ [for a 
recent review see, e.g., ref.~\cite{Dobbs}]. One also 
gets the correct order of magnitude of the superconducting 
and superfluid transition temperature $T_c$ when the model 
parameters are inferred from experiments in the normal state 
of the above systems. However, in Ref.\cite{PM1} it was found 
that when the model was treated nonperturbatively 
and one approached the border of magnetic long-range order, 
the quasiparticle spectrum showed qualitative changes not 
captured by the Eliashberg approximation. In Ref.\cite{PM1},
we raised the possibility that these qualitative changes,
namely the opening of a pseudogap in the quasiparticle
spectrum, were intrinsically nonperturbative in nature.
In this paper, we examine this possibility by comparing
the nonperturbative results to various kinds of 
perturbation-theoretic approximations.

The paper is organized as follows. In the next section 
we describe the model as well as the various 
perturbation-theoretic approximations to be compared 
to the Monte Carlo calculations. Section III contains 
the results of the nonperturbative and diagrammatic 
calculations. Section IV contains a discussion of the 
results and finally we give a summary and outlook

\section{Model}

The model and its motivation have been extensively discussed in
Ref.\cite{PM1}. Here we only give the definitions relevant to the
present discussion. We consider particles on a two-dimensional
square lattice whose Hamiltonian in the absence of interactions
is

\begin{equation}
{\hat h}_0(\tau) = -\sum_{i,j,\alpha} 
t_{ij}\psi^\dagger_{i\alpha}(\tau)\psi_{j\alpha}(\tau)
- \mu \sum_{i\alpha}\psi^\dagger_{i\alpha}(\tau)\psi_{i\alpha}(\tau)
\label{ham0}
\end{equation}

\noindent where $t_{ij}$ is the tight-binding hopping matrix,
$\mu$ the chemical potential and $\psi^\dagger_{i\alpha}$,
$\psi_{i\alpha}$ respectively create and annihilate a fermion 
of spin orientation $\alpha$ at site $i$. We take $t_{ij} = t$ 
if sites $i$ and $j$ are nearest neighbors and $t_{ij} = t'$ 
if sites $i$ and $j$ are next-nearest neighbors. 

To introduce interactions between the particles, we couple 
them to a dynamical molecular (or Hubbard-Stratonovich) 
field. It is instructive to consider two different types
of molecular fields. In the first instance, we consider a
vector Hubbard-Stratonovich field that couples locally to
the fermion spin density. We also consider the case
of a scalar field that couples locally to the fermion
number density. This case corresponds to a coupling to
charge-fluctuations or, within the approximation we are using 
here, "Ising"-like magnetic fluctuations where only longitudinal 
modes are present. The Hamiltonians at imaginary time $\tau$ 
for particles coupled to the fluctuating exchange or scalar 
dynamical field are then

\begin{eqnarray}
{\hat h}(\tau) & = & {\hat h}_0(\tau) - {g\over \sqrt{3}}\sum_{i\alpha\gamma}
{\bf M}_i(\tau)\cdot\psi^\dagger_{i\alpha}(\tau){\bf \sigma}_{\alpha\gamma}
\psi_{i\gamma}(\tau) 
\label{ham1} \\
{\hat h}(\tau) & = & {\hat h}_0(\tau) - g\sum_{i\alpha}
\Phi_i(\tau)\psi^\dagger_{i\alpha}(\tau)\psi_{i\alpha}(\tau)
\label{ham2}
\end{eqnarray}

\noindent where ${\bf M}_i(\tau) = (M^x_i(\tau),M^y_i(\tau),M^z_i(\tau))^T$
and $\Phi_i(\tau)$ are the real vector exchange and scalar 
Hubbard-Stratonovich fields respectively, and $g$ the 
coupling constant. The reason for the choice of an extra 
factor $1/\sqrt{3}$ in Eq.~(\ref{ham1}) becomes clear later.

Since we ignore the self-interactions of the molecular 
fields, their distribution is Gaussian and given by\cite{PhilMag,PM1}

\begin{eqnarray}
{\cal P}[{\bf M}] = {1\over Z} \exp\Bigg(-\sum_{{\bf q},\nu_n} 
{{\bf M}({\bf q},i\nu_n)\cdot {\bf M}(-{\bf q},-i\nu_n)
\over 2\alpha({\bf q},i\nu_n)}\Bigg)
\label{ProbM} \\
Z = \int D{\bf M}\exp\Bigg(-\sum_{{\bf q},\nu_n} 
{{\bf M}({\bf q},i\nu_n)\cdot {\bf M}(-{\bf q},-i\nu_n)
\over 2\alpha({\bf q},i\nu_n)}\Bigg)
\label{NormM}
\end{eqnarray}

\noindent in the case of a vector exchange molecular field and

\begin{eqnarray}
{\cal P}[\Phi] = {1\over Z}\exp\Bigg(-\sum_{{\bf q},\nu_n} 
{\Phi({\bf q},i\nu_n)\Phi(-{\bf q},-i\nu_n)
\over 2\alpha({\bf q},i\nu_n)}\Bigg)
\label{ProbPhi} \\
Z = \int D\Phi\exp\Bigg(-\sum_{{\bf q},\nu_n} 
{\Phi({\bf q},i\nu_n)\Phi(-{\bf q},-i\nu_n)
\over 2\alpha({\bf q},i\nu_n)}\Bigg)
\label{NormPhi}
\end{eqnarray}

\noindent in the case of a scalar Hubbard-Stratonovich field. 
In both cases $\nu_n = 2\pi n T$ since the dynamical molecular
fields are periodic functions in the interval $[0,\beta=1/T]$.
The Fourier transforms of the molecular fields are defined as

\begin{eqnarray}
{\bf M}_{\bf R}(\tau) & = & \sum_{{\bf q},\nu_n} {\bf M}({\bf q},i\nu_n) 
\exp\Big(-i[{\bf q}\cdot{\bf R}-\nu_n\tau]\Big)
\label{FourierM} \\
\Phi_{\bf R}(\tau) & = & \sum_{{\bf q},\nu_n} \Phi({\bf q},i\nu_n)
\exp\Big(-i[{\bf q}\cdot{\bf R}-\nu_n\tau]\Big)
\label{FourierP}
\end{eqnarray}

We consider the case where there is no long-range 
magnetic or charge order. The average of the dynamical
molecular fields must then vanish and their Gaussian
distributions Eqs.~(\ref{ProbM},\ref{ProbPhi}) are
completely determined by their variance 
$\alpha({\bf q},i\nu_n)$, which we take to be

\begin{equation}
\alpha({\bf q},i\nu_n) = \cases{{1\over 2}{T\over N}\chi({\bf q},i\nu_n)
&if ${\bf M}({\bf q},i\nu_n)$ or $\Phi({\bf q},i\nu_n)$ \cr &  is complex \cr
{T\over N}\chi({\bf q},i\nu_n)
&if ${\bf M}({\bf q},i\nu_n)$ or $\Phi({\bf q},i\nu_n)$ \cr &  is real}
\label{alpha}
\end{equation}

\noindent where $N$ is the number of allowed wavevectors
in the Brillouin zone. Then

\begin{eqnarray}
\Big<M_i({\bf q},i\nu_n)M_j({\bf k},i\Omega_n)\Big> & = &
{T\over N}\chi({\bf q},i\nu_n) \nonumber \\ & \times &
\delta_{{\bf q},-{\bf k}}
\delta_{\nu_n,-\Omega_n}\delta_{i,j}
\label{Mcorr} \\
\Big<\Phi({\bf q},i\nu_n)\Phi({\bf k},i\Omega_n)\Big> & = &
{T\over N}\chi({\bf q},i\nu_n)\delta_{{\bf q},-{\bf k}}\nonumber \\
& \times & \delta_{\nu_n,-\Omega_n}
\label{Pcorr}
\end{eqnarray}

\noindent where $<\dots>$ denotes an average over the 
probability distributions Eq.~(\ref{ProbM}) and 
Eq.~(\ref{ProbPhi}) for the vector and scalar cases
respectively. In order to compare the scalar and vector 
molecular fields, we take the same form for their
correlation function $\chi({\bf q},i\nu_n)$
and parametrize it as in Refs.\cite{ML1,ML2}. 
In what follows, we set the lattice spacing $a$ to unity.
For real frequencies, we have

\begin{equation}
\chi({\bf q},\omega) = {\chi_0\kappa_0^2\over \kappa^2 + \widehat{q}^2 
- i{\omega\over \eta(\widehat{q})}}
\label{chiML}
\end{equation}

\noindent where $\kappa$ and $\kappa_0$ are the correlation 
wavevectors or inverse correlation lengths in units of 
the lattice spacing, with and without strong 
correlations, respectively.  Let

\begin{equation}
\widehat{q}_{\pm}^2 = 4 \pm 2(\cos(q_x)+\cos(q_y)) 
\label{qdef}
\end{equation}

We consider commensurate charge fluctuations and
antiferromagnetic spin fluctuations, in which case
the parameters $\widehat{q}^2$ and $\eta(\widehat{q})$ 
in Eq.~(\ref{chiML}) are defined as

\begin{eqnarray}
\widehat{q}^2 & = & \widehat{q}_{+}^2 \\
\eta(\widehat{q}) & = & T_0\widehat{q}_{-}
\label{antiferro}
\end{eqnarray}

\noindent where $T_0$ is a characteristic temperature.  

We also consider the case of ferromagnetic 
spin-fluctuations, where the parameters 
$\widehat{q}^2$ and $\eta(\widehat{q})$ 
in Eq.~(\ref{chiML}) are given by

\begin{eqnarray}
\widehat{q}^2 & = & \widehat{q}_{-}^2 \\
\eta(\widehat{q}) & = & T_0\widehat{q}_{-}
\label{ferro}
\end{eqnarray}

\noindent $\chi({\bf q},i\nu_n)$ is related to the imaginary 
part of the response function $Im\chi({\bf q},\omega)$, 
Eq.~(\ref{chiML}), via the spectral representation

\begin{equation}
\chi({\bf q},i\nu_n) = -\int_{-\infty}^{+\infty}{d\omega\over \pi}
{Im\chi({\bf q},\omega)\over i\nu_n - \omega}
\label{chi_mats}
\end{equation}

\noindent To get $\chi({\bf q},i\nu_n)$ to decay as $1/\nu_n^2$ as
$\nu_n \rightarrow \infty$, as it should, we introduce a cutoff 
$\omega_0$ and take $Im\chi({\bf q},\omega) = 0$ for $\omega 
\geq \omega_0$. A natural choice for the cutoff is $\omega_0 
= \eta(\widehat{q})\kappa_0^2$.

In our model, the single particle Green's function is the 
average over the probability distributions ${\cal P}[{\bf M}]$ 
(Eq.~(\ref{ProbM})) or ${\cal P}[\Phi]$ (Eq.~(\ref{ProbPhi})) 
of the fermion Green's function in a dynamical vector or scalar 
field.

\begin{eqnarray}
{\cal G}(i\sigma\tau ; j\sigma'\tau') & = & 
\int D{\bf M}\; {\cal P}[{\bf M}]\; G(i\sigma\tau ; j\sigma'\tau'|[{\bf M}])
\label{gM} \\
{\cal G}(i\sigma\tau ; j\sigma'\tau') & = & 
\int D\Phi\; {\cal P}[\Phi]\; G(i\sigma\tau ; j\sigma'\tau'|[\Phi])
\label{gPhi}
\end{eqnarray}

\noindent where

\begin{equation}
G(i\sigma\tau ; j\sigma'\tau'|[{\bf M}]\;or \;[\Phi]) 
= -\big<T_\tau\{\psi_{i\sigma}(\tau)
\psi^\dagger_{j\sigma'}(\tau')\}\big> 
\label{gfield} 
\end{equation}

\noindent is the single particle Green's function in a
dynamical molecular field and is discussed at length in 
Ref.\cite{PM1}. In evaluating expressions 
Eqs.~(\ref{gM},\ref{gPhi}) one is summing over all Feynman 
diagrams corresponding to spin or charge-fluctuation 
exchanges\cite{PM1,PhilMag,Coleman}. The diagrammatic expansion 
of the Green's function, Eq.~(\ref{gM}) is shown pictorially in Fig. 1.
Since in our model no virtual fermion loops are present, there 
is no fermion sign problem\cite{PhilMag}.

In this paper we compare the results of the Monte Carlo 
simulations to various diagrammatic approximations for the same 
model. We denote by ${\cal G}_0({\bf p},i\omega_n)$ and
${\cal G}({\bf p},i\omega_n)$ the bare and dressed quasiparticle 
propagators respectively. They are given by

\begin{eqnarray}
{\cal G}_0({\bf p},i\omega_n) & = & {1\over i\omega_n 
- (\epsilon_{\bf p}-\mu)} \label{gBare} \\
{\cal G}({\bf p},i\omega_n) & = & {1\over i\omega_n 
- (\epsilon_{\bf p}-\mu) - \Sigma({\bf p},i\omega_n)} 
\label{gDressed}
\end{eqnarray}

\noindent where $\Sigma({\bf p},i\omega_n)$ is the quasiparticle
self-energy and $\epsilon_{\bf p}$ the tight-binding dispersion 
relation obtained from Fourier transforming the hopping matrix 
$t_{ij}$ in Eq.~(\ref{ham0}) and $\mu$ the chemical potential.
We consider four approximations to the quasiparticle self-energy
$\Sigma({\bf p},i\omega_n)$ whose diagrammatic representations
are shown in Fig. 2.

In Fig. 2a, the self-energy is approximated by first order 
perturbation theory in the exchange of magnetic or charge 
fluctuations and denoted $\Sigma^{1pt}({\bf p},i\omega_n)$. 
Fig. 2b shows the Eliashberg approximation in which the 
self-energy denoted $\Sigma^{1sc}({\bf p},i\omega_n)$ is given
by the first order self-consistent (or Brillouin-Wigner)
perturbation theory. The expressions for 
$\Sigma^{1pt}({\bf p},i\omega_n)$ or 
$\Sigma^{1sc}({\bf p},i\omega_n)$ in the case of quasiparticles 
coupled to magnetic or charge fluctuations are identical
(this is the reason for our choice of the factor $1/\sqrt{3}$
in Eq.~(\ref{ham1})) and given by

\begin{eqnarray}
\Sigma^{1pt}({\bf p},i\omega_n) & = &  g^2{T\over N}
\sum_{{\bf k},\Omega_n} \chi({\bf p}-{\bf k},i\omega_n-i\Omega_n)
\nonumber \\ & \times & {\cal G}_0({\bf k},i\Omega_n) \label{1pt} \\
\Sigma^{1sc}({\bf p},i\omega_n) & = &  g^2{T\over N}
\sum_{{\bf k},\Omega_n} \chi({\bf p}-{\bf k},i\omega_n-i\Omega_n)
\nonumber \\ & \times & {\cal G}({\bf k},i\Omega_n) \label{1sc}
\end{eqnarray}

\noindent where ${\cal G}_0({\bf k},i\Omega_n)$ and 
${\cal G}({\bf k},i\Omega_n)$ are the bare and dressed 
quasiparticle Green's functions defined in Eq.~(\ref{gBare}) 
and Eq.~(\ref{gDressed}) respectively. Fig. 2c shows the 
diagrammatic expansion corresponding to second order perturbation 
theory and we denote the self-energy corresponding to that 
approximation $\Sigma^{2pt}({\bf p},i\omega_n)$. The second order 
self-consistent approximation to the quasiparticle self-energy, 
denoted $\Sigma^{2sc}({\bf p},i\omega_n)$, is shown diagrammatically
in Fig. 2d. The expressions for $\Sigma^{2pt}({\bf p},i\omega_n)$
and $\Sigma^{2sc}({\bf p},i\omega_n)$ now depend on whether
the quasiparticles are coupled to the vector Hubbard-Stratonovich
field (magnetic fluctuations)  or scalar Hubbard-Stratonovich
field (charge fluctuations), because vertex corrections in the
two cases do not have the same coefficient or even the same sign.
The expressions for $\Sigma^{2pt}({\bf p},i\omega_n)$
and $\Sigma^{2sc}({\bf p},i\omega_n)$ for quasiparticles coupled
to magnetic fluctuations are given by

\begin{eqnarray}
\Sigma^{2pt}({\bf p},i\omega_n) & = &  g^2{T\over N}
\sum_{{\bf k},\Omega_n} \chi({\bf p}-{\bf k},i\omega_n-i\Omega_n)
\nonumber \\ & \times & {\cal G}_0({\bf k},i\Omega_n) \nonumber \\
& + & g^2{T\over N}\sum_{{\bf k},\Omega_n} 
\chi({\bf p}-{\bf k},i\omega_n-i\Omega_n)
\nonumber \\ & \times & {\cal G}_0({\bf k},i\Omega_n)
\Sigma^{1pt}({\bf p},i\omega_n){\cal G}_0({\bf k},i\Omega_n) 
\nonumber \\
& - & {1\over 3} \Bigg(g^2{T\over N}\Bigg)^2\sum_{{\bf k},\Omega_n}
\sum_{{\bf k}',\Omega_n'} \chi({\bf p}-{\bf k},i\omega_n-i\Omega_n)
\nonumber \\ & \times & {\cal G}_0({\bf k},i\Omega_n)
{\cal G}_0({\bf k}',i\Omega_n')
\chi({\bf k}-{\bf k}',i\Omega_n-i\Omega_n') \nonumber \\
& \times &{\cal G}_0({\bf p}-{\bf k}+{\bf k}',i\omega_n-i\Omega_n+i\Omega_n')
\label{2ptM} \\
\Sigma^{2sc}({\bf p},i\omega_n) & = &  g^2{T\over N}
\sum_{{\bf k},\Omega_n} \chi({\bf p}-{\bf k},i\omega_n-i\Omega_n)
\nonumber \\ & \times & {\cal G}({\bf k},i\Omega_n) \nonumber \\
& - & {1\over 3} \Bigg(g^2{T\over N}\Bigg)^2\sum_{{\bf k},\Omega_n}
\sum_{{\bf k}',\Omega_n'} \chi({\bf p}-{\bf k},i\omega_n-i\Omega_n)
\nonumber \\ & \times & {\cal G}({\bf k},i\Omega_n){\cal G}({\bf k}',i\Omega_n')
\chi({\bf k}-{\bf k}',i\Omega_n-i\Omega_n') \nonumber \\
& \times &{\cal G}({\bf p}-{\bf k}+{\bf k}',i\omega_n-i\Omega_n+i\Omega_n')
\label{2scM}
\end{eqnarray}

\noindent In the case of the scalar Hubbard-Stratonovich field, 
or coupling to charge fluctuations, the corresponding expressions
are

\begin{eqnarray}
\Sigma^{2pt}({\bf p},i\omega_n) & = &  g^2{T\over N}
\sum_{{\bf k},\Omega_n} \chi({\bf p}-{\bf k},i\omega_n-i\Omega_n)
{\cal G}_0({\bf k},i\Omega_n) \nonumber \\
& + & g^2{T\over N}\sum_{{\bf k},\Omega_n} 
\chi({\bf p}-{\bf k},i\omega_n-i\Omega_n){\cal G}_0({\bf k},i\Omega_n)
\nonumber \\ & \times &\Sigma^{1pt}({\bf p},i\omega_n)
{\cal G}_0({\bf k},i\Omega_n) \nonumber \\
& + & \Bigg(g^2{T\over N}\Bigg)^2\sum_{{\bf k},\Omega_n}
\sum_{{\bf k}',\Omega_n'} \chi({\bf p}-{\bf k},i\omega_n-i\Omega_n)
\nonumber \\ & \times & {\cal G}_0({\bf k},i\Omega_n)
{\cal G}_0({\bf k}',i\Omega_n')
\chi({\bf k}-{\bf k}',i\Omega_n-i\Omega_n') \nonumber \\
& \times &{\cal G}_0({\bf p}-{\bf k}+{\bf k}',i\omega_n-i\Omega_n+i\Omega_n')
\label{2ptC} \\
\Sigma^{2sc}({\bf p},i\omega_n) & = &  g^2{T\over N}
\sum_{{\bf k},\Omega_n} \chi({\bf p}-{\bf k},i\omega_n-i\Omega_n)
{\cal G}({\bf k},i\Omega_n) \nonumber \\
& + & \Bigg(g^2{T\over N}\Bigg)^2\sum_{{\bf k},\Omega_n}
\sum_{{\bf k}',\Omega_n'} \chi({\bf p}-{\bf k},i\omega_n-i\Omega_n)
\nonumber \\ & \times & {\cal G}({\bf k},i\Omega_n){\cal G}({\bf k}',i\Omega_n')
\chi({\bf k}-{\bf k}',i\Omega_n-i\Omega_n') \nonumber \\
& \times &{\cal G}({\bf p}-{\bf k}+{\bf k}',i\omega_n-i\Omega_n+i\Omega_n')
\label{2scC}
\end{eqnarray}

\noindent In Eqs.~(\ref{2ptM},\ref{2scM},\ref{2ptC},\ref{2scC}), 
${\cal G}_0({\bf k},i\Omega_n)$ and 
${\cal G}({\bf k},i\Omega_n)$ are the bare and dressed 
quasiparticle Green's functions defined in Eq.~(\ref{gBare}) 
and Eq.~(\ref{gDressed}) respectively.

\subsection{Mass renormalization parameter}

The strength of the coupling to the magnetic
or charge flucutations can be parametrized by a dimensionless
mass renormalization parameter $\lambda_Z$, which is defined 
as

\begin{eqnarray}
\lambda_Z & = & { \int_{-\infty}^{+\infty} {d\omega\over \pi}
<{1\over \omega}g^2 Im\chi({\bf p}-{\bf p'},\omega)>_{FS({\bf p},{\bf p'})} 
\over <1>_{FS({\bf p})} } \label{lambda1}
\end{eqnarray}

\noindent The Fermi surface averages are given by

\begin{eqnarray}
<\cdots>_{FS({\bf p})}  & = & \int {d^dp\over (2\pi)^d} \cdots 
\delta(\epsilon_{\bf p} - \mu) \label{FSaverage1} \\
<\cdots>_{FS({\bf p},{\bf p'})}  & = & \int {d^dp\over (2\pi)^d} 
{d^dp'\over (2\pi)^d}\cdots \nonumber \\ & \times &
\delta(\epsilon_{\bf p} - \mu) 
\delta(\epsilon_{\bf p'} - \mu)  \label{FSaverage2}
\end{eqnarray}

\noindent In practice, we compute the Fermi surface average with a 
discrete set of momenta and we replace the delta function by a finite
temperature expression

\begin{eqnarray}
\int {d^dp\over (2\pi)^d} & \longrightarrow &{1\over N}\sum_{\bf p} \\
\delta(\epsilon_{\bf p} - \mu) & \longrightarrow & {1\over T} f_{\bf p}(1-f_
{\bf p})
\label{FSaverageNum}
\end{eqnarray}

\noindent where $f_{\bf p}$ is the Fermi function. Note that 
${1\over T} f_{\bf p}(1-f_{\bf p}) \rightarrow 
\delta(\epsilon_{\bf p} - \mu)$ as $T \rightarrow 0$. We have used 
$T = 0.1t$ and $N = 128^2$ in all of our calculations. The finite 
temperature effectively means that van Hove singularities will be 
smeared out.

Note that the Fermi surface average that appears in $\lambda_Z$, 
Eq.~(\ref{lambda1}) plays a role similar to that of 
$\alpha^2F(\omega)/\omega$ in the case of phonon mediated 
superconductivity. One therefore expects $\lambda_Z\sim 1$
to indicate the crossover between weak and strong coupling.

\section{Results}

The quasiparticle dispersion relation for the 
two-dimensional square lattice is obtained from 
Eq.~(\ref{ham0}). We measure all energies and temperatures 
in units of the nearest-neighbor hopping parameter $t$. 
We set the next-nearest-neighbor hopping parameter 
$t'=-0.45t$. The chemical potential is adjusted so that
the electronic band filling is $n=0.9$. The dimensionless 
parameters describing the molecular field correlations
are $g^2\chi_0/t$, $T_0/t$, $\kappa_0$ and $\kappa$. 
We chose a representative value for $\kappa_0^2 = 12$, 
and set $T_0 = 0.67t$ as in the earlier work\cite{PM1}. 
For an electronic bandwidth of $1eV$, $T_0\approx 1000^\circ$K.
We only consider one value of the coupling constant
$g^2\chi_0/t = 2$. In the random phase approximation,
the magnetic instability would be obtained for a value
of $g^2\chi_0/t$ of the order of 10. We consider what 
happens to the quasiparticle spectrum at a fixed 
temperature $T = 0.25t$ as the inverse correlation 
length $\kappa$ changes, as in Ref.\cite{PM1}.

All the calculations were done on a 8 by 8 spatial lattice.
In the Monte Carlo calculations we used 41 imaginary time 
slices, or equivalently 41 Matsubara frequencies for the 
molecular fields, ${\bf M}({\bf q},i\nu_n)$ and 
$\Phi({\bf q},i\nu_n)$ 
($\nu_n = 2\pi n T$, with $ n=0,\pm 1,\dots,\pm 20$).
In the diagrammatic calculations, we used between 40
to 60 fermion Matsubara frequencies.

By analytic continuation of the single particle Green's 
function ${\cal G}({\bf k},\tau)$  one can obtain the 
quasiparticle spectral function $A({\bf k},\omega) 
= - {1\over \pi}Im\;{\cal G}_R({\bf k},\omega)$ and 
the tunneling density of states $N(\omega) = {1\over N}
\sum_{\bf k}A({\bf k},\omega)$, where
${\cal G}_R({\bf k},\omega)$ is the retarded single
particle Green's function. The imaginary time Monte Carlo 
data is analytically continued with the Maximum Entropy 
method\cite{MaxEnt}, using the same methodology as in the 
earlier work\cite{PM1}. We used 10000 Monte Carlo samples
grouped into 100 bins of 100 samples each. We always use
a flat default model in the Maximum Entropy calculations.
To provide a fair comparison between diagrammatic and 
nonperturbative calculations, one should use the same 
analytic continuation method (with the same parameters) in all cases. 
Therefore, we generated 100 noisy measurements by adding 
Gaussian random noise to the results of the diagrammatic calculations 
and analytically continued ${\cal G}({\bf k},\tau) = {\cal G}_0({\bf k},\tau)
+ T\sum_{\omega_n}e^{-i\omega_n\tau}\big({\cal G}({\bf k},i\omega_n)-
{\cal G}_0({\bf k},i\omega_n)\big)$ using the Maximum
Entropy method as well, with the same default model as in
the corresponding analytic continuation of the Monte Carlo
data. The scheme is not perfect, however. While the variance 
of the Gaussian noise added to the diagrammatic Green's functions 
was chosen such that the statistical uncertainty of the average over
the 100 noisy samples was identical to that in the corresponding Monte
Carlo Green's function, the correlations in the errors for 
different values of $\tau$ present in the Monte Carlo results 
cannot be easily modeled. The Gaussian random numbers added to 
the diagrammatic Green's function were therefore taken to be
independent of each other, and thus the noise in the diagrammatic
and Monte Carlo Green's functions did not have identical
statistical properties. In spite of this, the present scheme 
is almost certainly better than the alternatives.

\subsection{Antiferromagnetic spin-fluctuations}

Figs. 3,4 and 5 show the comparison, for different values of 
$\kappa^2$, between the nonperturbative calculations of
the quasiparticle Green's function ${\cal G}({\bf k},\tau)$,
spectral function $A({\bf k},\omega)$ 
and tunneling density of states $N(\omega)$ and those obtained 
from the approximations $\Sigma^{1pt}({\bf p},i\omega_n)$, 
$\Sigma^{2pt}({\bf p},i\omega_n)$, $\Sigma^{1sc}({\bf p},i\omega_n)$, 
and $\Sigma^{2sc}({\bf p},i\omega_n)$ to the quasiparticle 
self-energy.

Fig. 3 shows our results for $\kappa^2 = 24$. For this value of the
inverse correlation length squared, the mass renormalization parameter
$\lambda_Z \approx 0.05$. The coupling to the antiferromagnetic
spin-fluctuations is therefore weak. Not surprisingly, the 
quasiparticle Green's function, spectral function and tunneling
density of states obtained from the various diagrammatic approximation
agree well with the Monte Carlo results. At ${\bf k} = (\pi,0)$,
the difference between the nonperturbative Green's function 
${\cal G}(({\bf k},\tau)$ and its diagrammatic approximations
is of the order of 0.001t for all values of $\tau$. There is virtually
no difference between the straightforward perturbation-theoretic
calculations of the spectral function and their self-consistent 
counterparts, in first and second order, which is expected for weak 
coupling. Thus the small difference in the spectral functions 
$A({\bf k},\omega)$ at ${\bf k} = (\pi,0)$, 
seen in Fig 3b, to the extent that they are not an artifact of the 
analytic continuation, must come from the vertex corrections. Since 
the first order spectral functions are slightly sharper than the 
second order ones, the first order vertex corrections result in an 
increased spin-fluctuation interaction, as pointed out in 
Refs.\cite{Vertex1,Vertex2,PM1}. The Monte Carlo spectral function is also 
somewhat broader than the diagrammatic calculations, and provided 
again that it is not an artifact of the analytic continuation, this 
suggests that the higher order diagrams lead to a further increase 
of the spin-fluctuation interaction.

The results for $\kappa^2 = 4$ are shown in Fig. 4. This value
of $\kappa^2$ gives a mass renormalization parameter 
$\lambda_Z \approx 0.45$. One is now in the intermediate coupling
regime. The quasiparticle Green's function ${\cal G}({\bf k},\tau)$ 
and spectral function $A({\bf k},\omega)$ at ${\bf k} = (\pi,0)$ as 
well as the tunneling density of states obtained from the various 
diagrammatic approximations agree qualitatively with the Monte Carlo 
results. There are, however, noticeable quantitative differences, 
not surprisingly much more so than for $\kappa^2 = 24$. The largest
difference between the Green's functions obtained from the diagrammatic
approximations and the nonperturbative calculations is now bigger
than the width of the lines and is roughly an order of magnitude (0.01t)
larger than for $\kappa^2 = 24$, which is not unexpected since the
mass renormalization parameter is also about an order of magnitude
greater for $\kappa^2 = 4$ than for $\kappa^2 = 24$.  With the
above caveat regarding the analytic continuation, one can make a few 
additional remarks. First of all, there is now a difference between the 
straightforward perturbation-theoretic results and the self-consistent 
calculations of the spectral function $A({\bf k},\omega)$ 
at ${\bf k} = (\pi,0)$, both at first and
second order. In particular, the second order self-consistent
spectral function is slightly broader than the first order
self-consistent one, an indication that the first order vertex 
correction leads to an enhancement of the effective spin-fluctuation
interaction, in agreement with Refs.\cite{Vertex1,Vertex2,PM1}. The
nonperturbative $A({\bf k}=(\pi,0),\omega)$ is broader than the
second order self-consistent result, which would imply the
higher order vertex corrections are further enhancing the
magnetic interaction. Note, that the second order
perturbation-theoretic $A({\bf k}=(\pi,0),\omega)$ is slightly 
broader than its self-consistent counterpart (the dressing of Green's
functions tends to reduce the effect of interactions) and agrees 
very well with the Monte Carlo result. This may be due
to a cancellation of errors (or the analytic continuation procedure)
since the agreement between the nonperturbative tunneling density of 
states $N(\omega)$ and the second order perturbation-theoretic 
$N(\omega)$ is not as good.

As $\kappa^2 \approx 1$, the quasiparticle mean free
path becomes of the order of the magnetic correlation
length for some wavevectors near the Fermi surface, 
the quasiparticles then can't tell there is
no long-range order, and this marks the onset of
pseudogap behavior\cite{PM1}. For $\kappa^2 = 1$,
the mass renormalization parameter is
$\lambda_Z \approx 1.1$. One is therefore in the
strong coupling regime. The results of our calculations
for $\kappa^2 = 1$ are shown in Fig. 5. The developing 
pseudogap in the spectral function $A({\bf k},\omega)$ at 
${\bf k} = (\pi,0)$ (Fig. 5b) and in the tunneling
density of states $N(\omega)$ (Fig. 5e) found in the 
nonperturbative Monte Carlo calculations is not seen 
in any of the diagrammatic approximations considered here, 
which therefore fail qualitatively. Given that one is
in the strong coupling regime $\lambda_Z > 1$, the
breakdown of perturbation theory should not come as
a surprise. The maximum difference in the quasiparticle
Green's function ${\cal G}({\bf k},\tau)$ between the
nonperturbative and diagrammatic calculations is now
of the order 0.1t, and hence an order of magnitude larger 
than for $\kappa^2 = 4$ and a couple of orders of magnitude
larger than in the weak coupling regime with $\kappa^2 = 24$.
It is therefore not suprprising that the quasiparticle spectra
that give rise to these rather different imaginary time
Green's functions turn out to show qualitative differences. 
Note that for $\kappa^2 = 1$, there is nearly as 
much difference between the perturbation-theoretic and 
self-consistent approximations of the same order as 
there are between calculations of the same type at first 
and second order.

\subsection{Ferromagnetic spin-fluctuations}

Figs. 6-9 show our results for the quasiparticle
Green's function ${\cal G}({\bf k},\tau)$,
spectral function $A({\bf k},\omega)$ and tunneling 
density of states $N(\omega)$ for several values of
$\kappa^2$. 

We start with $\kappa^2 = 24$, for which
the mass renormalization parameter $\lambda_Z \approx 0.05$
for coupling to ferromagnetic spin-flucutations.
In this weak coupling regime, Fig. 6 shows that the
results of the various diagrammatic calculations are
in good agreement with the Monte Carlo results. As in
the corresponding antiferromagnetic case, at ${\bf k} = (\pi,0)$,
the difference between the nonperturbative Green's function 
${\cal G}(({\bf k},\tau)$ and its diagrammatic approximations
is of the order of 0.001t for all values of $\tau$. Moreover, 
there is virtually no difference between the straightforward 
perturbation-theoretic calculations of the spectral 
function and their self-consistent counterparts, in first 
and second order. Thus the small difference in the spectral 
functions $A({\bf k},\omega)$ at ${\bf k} = (\pi,0)$, 
seen in Fig. 6b must come from the vertex corrections. Since 
the first order spectral functions are slightly sharper than the 
second order ones, the first order vertex corrections result in an 
increased spin-fluctuation interaction. Given the smallness of
the difference between the first and second order results and 
the ill-posed nature of the analytic continuation problem, one 
should take the above remark with some degree of caution. 
In Ref.\cite{PM1}, however, it was shown that the increase in 
the effective interaction induced by the first order vertex 
correction is due to the spin dependence of the interaction, 
and thus should occur for quasiparticles coupled to either 
antiferromagnetic or ferromagnetic fluctuations. Our 
analytically continued results are at least consistent with this.
From Fig. 6b, one also sees that the Monte Carlo spectral 
function is slightly broader than the first or second order
results, as in the corresponding antiferromagnetic case.
With the above caveat on the nature of the analytic continuation
problem, this would suggest the higher order diagrams not
included in our perturbation-theoretic approximations lead
to a further enhancement of the magnetic interaction, as
in the corresponding antiferromagnetic case.

For $\kappa^2 = 4$, the mass renormalization parameter
$\lambda_Z\approx 0.45$ and one is therefore in an intermediate
coupling regime. Fig. 7 shows that for this value of $\kappa^2$,
the diagrammatic approximations all qualitatively agree
with the Monte Carlo results. The quantitative agreement is,
not surprisingly, not as good as in the weak coupling limit with
$\kappa^2 = 24$. One notes a number of similarities between the
results of Fig. 6 and the corresponding antiferromagnetic
case, shown in Fig. 4: (i) the second order perturbation
theory results for $A({\bf k},\omega)$ give the best
agreement with the nonperturbative calculation, (ii)
since the spectral function in either second order 
calculation, which include vertex corrections, is slightly 
broader in $\omega$ than the corresponding first order result,
we conclude that first order vertex corrections lead
to an enhancement of the effective quasiparticle 
interaction, which is what is expected on the basis of the 
arguments made in Refs.\cite{Vertex1,Vertex2,PM1} (iii) the spectral 
function obtained by Monte Carlo sampling of the Gaussian 
dynamical molecular fields is slightly broader than the second 
order results, which to the extent this is not an artifact of 
the Maximum Entropy analytic continuation is an indication that 
higher order spin-fluctuation exchanges not included in the 
diagrammatic approximations considered lead to a further 
enhancement of the effective quasiparticle interaction. 

The dynamical exponent $z$ is larger for ferromagnetic than 
antiferromagnetic spin fluctuations. Hence the effective 
dimension $d+z=5$ in the ferromagnetic versus $d+z=4$ in 
the antiferromagnetic case and the standard theory 
of quantum critical phenomena\cite{HertzMillis} leads
one to expect weaker corrections for higher effective 
dimensions. The perturbative calculations qualitatively fail
at $\kappa^2 \approx 1$ in the antiferromagnetic case and 
on the basis of the above arguments one would expect that
the breakdown of perturbation theory in the case of 
ferromagnetic fluctuations, if it happens, would
occur for a smaller value of $\kappa^2$ or larger values
of the mass renormalization parameter $\lambda_Z$. Indeed, at 
$\kappa^2 = 1$, $\lambda_Z \approx 1.2$ and therefore
one is in the strong coupling regime. Fig. 8 shows that 
while for this value of $\kappa^2$ the diagrammatic calculations 
still agree qualitatively with the Monte Carlo results, 
unsurprisingly there are larger quantitative differences than 
in the case $\kappa^2 = 4$ shown in Fig. 7.

Our results for $\kappa^2 = 0.25$, for which the mass renormalization
parameter $\lambda_Z \approx 2.3$ are shown in Fig. 9. The spectral 
function $A({\bf k},\omega)$ obtained from the nonperturbative
Monte Carlo calculations shows a double peak structure. This has
been interpreted in Ref.\cite{PM1} as an effective spin-splitting 
of the quasiparticle spectrum induced by the local ferromagnetic
order. In looking at the evolution of the spectral function 
$A({\bf k},\omega)$ as $\kappa^2$ is decreased, one first sees a 
broadening of $A({\bf k},\omega)$ and then, the broad quasiparticle
peak splits into two. The Monte Carlo calculations show 
very little suppression of the quasiparticle spectral weight or
density of states between the two split peaks. A look at Figs. 7c,
8c, and 9e reveals that for $\kappa^2 \leq 4$, $N(\omega=0)t \approx 0.15$
and depends very little on $\kappa^2$. This is is sharp
contrast to the case of antiferromagnetic fluctuations discussed
in the previous section. This difference is to be expected of course,
since the antiferromagnetic state is gapped while the ferromagnetic
state is not. It is clear that none of the diagrammatic approximations
considered here reproduce this spin-splitting of the broad quasiparticle 
peak in $A({\bf k},\omega)$ and tunneling density of states $N(\omega)$ well.
In fact, the first order perturbation theoretic result shows a strong
suppression of the tunelling density of states, which clearly
doesn't describe the precursor to the ferromagnetic state well, 
and therefore can be considered to fail qualitatively. We observe that
the first order perturbation theoretic calculation failed to show a 
suppression of the tunneling density of states in the antiferromagnetic 
case where it is obtained in the nonperturbative calculations as 
expected (see previous subsection) but does show such a pseudogap
in the ferromagnetic case where it isn't expected and doesn't appear
in the nonperturbative calculations. It therefore qualitatively fails
in both cases. Another clear sign that not all is well with the
perturbation expansion is the large quantitative differences between 
the one-loop and two-loop results in Fig.9, something that could be 
expected at $\lambda_Z \approx 2.3$. In the view of the differences 
between the imaginary time Green's function ${\cal G}({\bf k},\tau)$ 
obtained from the Monte Carlo simulations and those of the various 
perturbation-theoretic approximations shown in Figs. 9a and 9c 
which are of the order of 0.1t, one would expect the spectral
functions that produce these rather different imaginary time
Green's functions to be rather different themselves.
\subsection{Charge fluctuations}

The results of our calculations of the quasiparticle 
Green's function ${\cal G}({\bf k},\tau)$,
spectral function $A({\bf k},\omega)$ and tunneling 
density of states $N(\omega)$ for several values of
$\kappa^2$ are shown in Figs. 10, 11 and 12. 

For the model studied here, the mass renormalization parameter 
$\lambda_Z$ is the same for charge and antiferromagnetic 
fluctuations. Therefore the results of the calculations for
$\kappa^2 = 24$ shown in Fig. 10 correspond to 
$\lambda_Z\approx 0.05$, namely the coupling to the charge
fluctuations is weak. The agreement between the Monte Carlo
results and those of the various diagrammatic approximations
is good. As seen in Fig. 10a, the difference between the
nonperturbative imaginary time Green's function and its
perturbative approximations at ${\bf k} = (\pi,0)$ is less 
than the width of the line and of the order of 0.001t for 
all imaginary times $\tau$. If one compares the results of the 
perturbation-theoretic calculations at first and second order,
one sees from Fig. 10b that there is virtually no difference
between the spectral functions $A({\bf k},\omega)$ obtained
by straightforward perturbation theory or the self-consistent
calculation at either first or second order. Hence the slight
difference bewteen the first and second order calculations,
to the extent they aren't an artifact of the analytic continuation,
must come from vertex corrections. In contrast to the case of
coupling to antiferromagnetic fluctuations, the spectral functions
at second order are slightly narrower than their first order
counterpart. This suggests the first order vertex correction
acts to reduce the effective charge fluctuation interaction,
in agreement with the arguments presented in Ref.\cite{PM1}.
Moreover, the nonperturbative $A({\bf k},\omega)$ at 
${\bf k} = (\pi,0)$ is slightly broader than the second order
results, which would indicate that the higher order diagrams
lead to an enhancement of the effective charge fluctuation
interaction, as in the case of a coupling to antiferromagnetic 
fluctuations. While this observation is made on the basis
of analytically continued results, it is consistent
with the results for other values of $\kappa^2$ presented
below, where the enhancement of the effective charge fluctuation 
mediated interaction by higher than second order diagrams can
be shown to occur on general grounds.

Fig. 11 shows the the quasiparticle Green's function 
${\cal G}({\bf k},\tau)$, spectral function $A({\bf k},\omega)$ 
at ${\bf k} = (\pi,0)$ and tunneling density of states $N(\omega)$ for 
$\kappa^2 = 4$, for which $\lambda_Z \approx 0.45$. For this value
of $\kappa^2$ corresponding to an intermediate coupling regime, 
the reader will notice that the results of the second order 
perturbation theory (self-consistent or not) are not displayed 
in the figures. The reason is that both second order approximations,
$\Sigma^{2pt}({\bf p},i\omega_n)$ and
$\Sigma^{2sc}({\bf p},i\omega_n)$, for the model parameters
considered here, effectively violate causality requirements
in that the Eliashberg renormalization factor 
$Z({\bf p},i\omega_n)$ becomes less than one. In terms of
the quasiparticle self-energy $\Sigma({\bf p},i\omega_n)$,
$Z({\bf p},i\omega_n) = 1 - {1\over \omega_n}Im\Sigma({\bf p},i\omega_n)
= 1 - \int_{-\infty}^{+\infty} {d\omega\over \pi}
{Im\Sigma_R({\bf p},\omega)\over \omega^2 + \omega_n^2}$
where $Im\Sigma_R({\bf p},\omega)$ is the imaginary part of 
the retarded self-energy and we have made use of the spectral 
representation for the self-energy $\Sigma({\bf p},i\omega_n)
=-\int_{-\infty}^{+\infty} {d\omega\over \pi}
{Im\Sigma_R({\bf p},\omega)\over i\omega_n - \omega}$. Causality
demands that the retarded Green's function be analytic in the 
upper-half complex frequency plane and therefore that the imaginary
part of the retarded self-energy be always less than or equal to
zero ($Im\Sigma_R({\bf p},\omega) \leq 0$) for all values of
${\bf p},\omega$. This in turn means that $Z({\bf p},i\omega_n)\geq 1$
for all values of ${\bf p},\omega$. One can write the second order
Eliashberg renormalization factor $Z^{(2)}({\bf p},i\omega_n) = 1
+ \Delta Z^{(1)}({\bf p},i\omega_n) + \Delta Z^{(2)}({\bf p},i\omega_n)$,
where $\Delta Z^{(i)}({\bf p},i\omega_n)$ is the change in 
$Z({\bf p},i\omega_n)$ coming from the $i^{th}$ order diagrams.
$\Delta Z^{(1)}({\bf p},i\omega_n)$ is always greater than zero 
and therefore poses no problem as far as the condition 
$Z({\bf p},i\omega_n)\geq 1$ is concerned. In the 
charge-fluctuation case, as was explained in Ref.\cite{PM1}, 
the first order vertex correction has the opposite sign compared
to the spin-fluctuation case, and leads to a suppression
of the effective quasiparticle interaction. The enhancement
of the quasiparticle spin-fluctuation vertex comes from the
transverse magnetic fluctuations that manage to overcome the
reduction of the effective coupling due to the longitudinal
fluctuations. Because of this cancellation effect, not only
is the sign of the first order vertex correction different
in the magnetic case, it is also smaller in magnitude than
in the charge-fluctuation case, under otherwise similar 
conditions, as can be seen from the factor 1/3 in 
Eqs.~(\ref{2ptM},\ref{2scM}) not present in the corresponding
charge-fluctuation case in Eqs.~(\ref{2ptC},\ref{2scC}).
The different sign of the vertex corrections in the charge
and magnetic cases means that while in the magnetic case
$\Delta Z^{(2)}({\bf p},i\omega_n)\geq 0$ and at second order
$Z^{(2)}({\bf p},i\omega_n)$ is  always $\geq 1$, in the charge
fluctuation case $\Delta Z^{(2)}({\bf p},i\omega_n)\leq 0$.
For $\kappa^2 \leq 4$, we find that the second order contribution 
to the Eliashberg renormalization factor is greater in magnitude 
than the first order contribution, $|\Delta Z^{(2)}({\bf p},i\omega_n)| > 
|\Delta Z^{(1)}({\bf p},i\omega_n)|$ and since it has the
opposite sign, $Z^{(2)}({\bf p},i\omega_n) = 1
+ \Delta Z^{(1)}({\bf p},i\omega_n) 
+ \Delta Z^{(2)}({\bf p},i\omega_n) \leq 1$. Note that the
nonperturbative calculations always satisfy 
$Z({\bf p},i\omega_n) \geq 1$, and the problem only arises
in the perturbative approximation and is a sign that, for
$\kappa^2 \leq 4$, the perturbation expansion for the charge-fluctuation 
case is quite badly behaved, possibly even more so than for 
magnetic fluctuations. Aso, the fact that the nonperturbative 
calculations always satisfy $Z({\bf p},i\omega_n) \geq 1$ is
a proof that the higher than second order diagrams contribute
to an enhancement of the charge-fluctuation interaction for
these values of $\kappa^2$. 

Fig. 11b shows that the spectral function 
$A({\bf k},\omega)$ obtained from the nonperturbative 
calculations is noticeably sharper than those produced by 
the first order self-consistent calculations. This means 
that for $g^2\chi_0/t = 2$ and $T_0 = 0.67t$ the first and 
higher order vertex corrections suppress the effective 
quasiparticle interaction. It is therefore not surprising 
that there are no qualitative differences between 
the nonperturbative and diagrammatic calculations. There
are quantitative differences, however, and these are more pronounced
than in the case of a coupling of quasiparticles to 
antiferromagnetic spin-fluctuations for the same value 
of $\lambda_Z$ shown in Fig. 4.

Finally, Fig. 12 shows our results for $\kappa^2 = 0.25$,
for which $\lambda_Z\approx 1.8$, hence in the strong
coupling regime. The difference between the nonperturbative
imaginary time Green's function and its first order
diagrammatic approximations seen in Figs. 12b and 12d
is a clear indication of the breakdown of perturbation
theory. But even in this strong coupling regime, a CDW-precursor
pseudogap in the spectral function $A({\bf k},\omega)$, which 
can be expected to occur on general 
grounds\cite{PM1} is not seen. The pseudogap effects 
in the charge fluctuation case thus require a stronger 
coupling still (larger coupling constant $g^2$ or 
smaller value of $\kappa^2$).

\section{Discussion}

In Ref.\cite{PM1}, we showed that the magnetic pseudogap induced
by a coupling to antiferromagnetic spin-fluctuations and the
spin-splitting of the quasiparticle peak induced by a coupling 
to ferromagnetic spin fluctuations were not captured by the
first order self-consistent, or Eliashberg, approximation. The
main result of this paper, is that these phenomena also lie 
beyond the two magnetic-fluctuation exchange theories 
(self-consistent or not), which contain first order vertex corrections.
While this does obviously not constitute a proof, these results
are consistent with the conjecture expressed in Ref.\cite{PM1}
that the pseudogap effects found in the Monte Carlo calculations
are intrinsically nonperturbative in nature. Since the 
calculations reported here show that the first order vertex 
corrections alone do not produce a magnetic pseudogap, the
physics of that state must then mainly come from the higher
order spin-fluctuation exchange processes. The results presented
here and in Ref.\cite{PM1} also indicate that a CDW pseudogap 
induced by coupling to the scalar dynamical molecular field 
(Eq.~(\ref{ham2})) must also originate from high order 
charge-fluctuation exchange processes. Close enough to a second 
order CDW transition, the diverging CDW 
correlation length is bound to exceed the characteristic length 
scale for quasiparticles and the calculations of Ref.\cite{MS} 
showed that when this happens a pseudogap opens in the quasiparticle 
spectrum. The first order vertex correction can't produce
the pseudogap state, since as we have seen, in the case of charge
fluctuations it leads to a suppression of the interaction. In fact
we even found that for the range of model parameters considered
here, the second order diagrams more than cancel the contribution
from the first order terms leading to a second order Eliashberg 
renormalization parameter $Z({\bf p},i\omega_n) \leq 1$, which
is inconsistent with causality requirements. Moreover, we expect
this "over-cancellation" effect to get worse as $\kappa^2$ gets 
smaller than the lowest value considered here, $\kappa^2 = 0.25$.
Since $Z({\bf p},i\omega_n)$ must be $\geq 1$ when all the diagrams
are summed up, as in the Monte Carlo simulations, one can conclude
that the higher than second order terms must give a contribution
$\Delta Z$ to $Z$ which is positive. Therefore, higher order 
charge-fluctuation exchange processes produce an enhancement of 
the effective quasiparticle interaction, as in the magnetic case, 
and it must be through this enhancement of the effective interaction 
that a pseudogap can appear in the quasiparticle spectrum on the 
border of long-range CDW order.

These observations lead one to a unified picture of the pseudogap
state found in our model of quasiparticles coupled to spin or 
charge fluctuations. When the dynamical molecular field correlation 
length exceeds the characteristic length scale for quasiparticles, 
either the thermal de Broglie wavelength\cite{MS,Vilk,Moukouri} 
or mean free path\cite{PM1}, the quasiparticles effectively see long-range 
order and this marks the onset of the pseudogap state. This state
must be produced by high order spin or charge-fluctuation exchanges
which contain subtle quantum mechanical coherence effects. In the
magnetic fluctuation case, the first order vertex correction 
favors the pseudogap state, while in the charge fluctuation 
case it suppresses it.  This implies one has to be closer to
the boundary of long-range charge order to observe a pseudogap
than one has to be to the boundary of magnetic long-range order,
under otherwise similar conditions. As the dynamical molecular
field correlation length increases, the mass renormalization
parameter $\lambda_Z$ gets larger, and therefore the many-body
effects become stronger. Our results show that the agreement
between the results of the Monte Carlo simulations and the
perturbation-theoretic results gets worse as $\lambda_Z$ increases,
and that not surprisingly, the perturbation-theoretic calculations
break down when one enters the strong coupling regime $\lambda_Z > 1$,
where the pseudogap is found. A more rigourous analysis of the 
relevance of the effective quasiparticle interactions as $\kappa^2$
increases or as the energy scales are decreased would require
a renormalization group (RG) treatment\cite{RG}. Recent RG 
calculations\cite{Pepin} on the border of the ferromagnetic state 
indicate that the quasiparticle interactions are indeed relevant 
in $d\leq 3$, and the RG flows to strong coupling as the energy
cutoff is decreased.

One would like to understand what property of the full vertex
function $\Gamma^i_{\alpha,\gamma}({\bf x},\tau;{\bf x}',\tau'|
{\bf x}",\tau") = <T_\tau\psi_\alpha({\bf x},\tau)
\psi^\dagger_\gamma({\bf x}',\tau')M^i({\bf x}",\tau")\}>$ is 
responsible for the appearance of the pseudogap and seems to
be missing in the first order approximation to 
$\Gamma^i_{\alpha,\gamma}({\bf x},\tau;{\bf x'},\tau'| {\bf x"},\tau")$.

Our physical picture of the pseudogap state emerging from quantum
mechanical coherence effects contained in high order Feynman 
diagrams is to be contrasted with the results of 
Refs.\cite{KS,Prelovsek,LAR} where a suppression of the 
quasiparticle tunneling density of states at the Fermi 
level is obtained in the single spin or charge-fluctuation 
exchange approximation. This effect is typically obtained with 
relatively large magnetic or charge correlation lengths. 
In our model, the calculations reported here and in 
Ref.\cite{PM1} show that as one approaches the 
border of magnetic long-range order, 
$\kappa^2 \rightarrow 0$, the multiple spin-fluctuation 
exchange processes become important long before a suppression
of the quasiparticle tunneling density of states at the Fermi
level is seen in the first order perturbation-theoretic and 
self-consistent calculations. Indeed, pseudogap effects are only
obtined in our calculations when the dimensionless mass 
renormalization parameter $\lambda_Z > 1$, i.e in the strong 
coupling regime where one doesn't expect diagrammatic perturbation
theory to give reliable approximations. The above finding is likely
to be valid more generally, since the intuitive arguments
for the physical origin of the pseudogap\cite{PM1,MS,Vilk,Moukouri}
lead one to expect the breakdown of Migdal's theorem to be a
generic occurence near a spin or charge instability. There is
also an important difference between a vertex correction
induced pseudogap and a single-fluctuation exchange pseudogap.
In the latter case, there is no essential distinction bewteen
spin and charge fluctuations, in that at the single-fluctuation
exchange level, for a given fluctuation spectrum the spin and 
charge-fluctuation theories of the quasiparticle spectral 
function can be made identical by an appropriate scaling of
the coupling constant to the molecular field. This is no 
longer the case when vertex corrections are included, since 
these actually depend on the nature of the Hubbard-Stratonovich
field, in our case vector versus scalar. The distinction could
turn out to be essential, since we find, for a range of model 
parameters, that a pseudogap is observed for quasiparticles 
coupled to spin fluctuations but not in the corresponding 
charge-fluctuation case.

Moukouri et al.\cite{Moukouri} have developed a many-body
theory of the precursor pseudogap to the Mott transition
in the half-filled Hubbard model. Their theory is inspired
by the fluctuation exchange approximation (FLEX)\cite{FLEX}
in which bare spin and charge susceptibilities are used to
build up the effective quasiparticle interaction, 
corresponding to $g^2\chi({\bf q},\omega)$ 
in our model. The key respect in which the theory of 
Moukouri et al.\cite{Moukouri} differs from FLEX is that 
the coupling to spin and charge fluctuations are not given
by the bare on-site Coulomb repulsion, but by renormalized 
parameters determined self-consistently in such a way that
an exact relationship between the single and two-particle
Green's functions is satisfied. This last step goes beyond 
perturbation theory and it is therefore plausible that the
precursor pseudogap to the Mott transition seen in the
Monte Carlo simulations of the half-filled Hubbard 
model\cite{Moukouri} is also nonperturbative in origin.
The analog of their scheme for the present model would be
the use of the first order perturbation theory approximation
for the quasiparticle self-energy $\Sigma^{(1pt)}({\bf p},\omega)$,
Eq.~(\ref{1pt}) and a simultaneous renormalization of the
coupling constant $g$ and the correlation wavevector $\kappa^2$.
A renormalization of the coupling constant $g$ could account 
for all vertex corrections provided they are
local in space and time. 

One can indeed get a pseudogap
in the tunneling density of states with the first order 
perturbation theory approximation to $\Sigma({\bf p},\omega)$ 
(Eq.~(\ref{1pt})), as in Refs.\cite{KS,Prelovsek}, provided $\kappa^2$ 
is renormalized to lower values and $g$ renormalized to higher ones.
One would thus have to renormalize the model to stronger coupling,
roughly to values of $\lambda_Z \sim 5$. It should be clear that
in this regime, first order perturbation theory is not controlled.
One would also naively expect a sensible renormalization 
scheme that goes beyond the one-loop level to lead 
to renormalized values of $\kappa^2$ larger than the bare 
value. The renormalized theory should be further away 
from the magnetic instability than the one-loop
approximation rather than closer to it, since ideally one
would like the improved theory to satisfy the Mermin-Wagner
theorem in two dimensions. If $\kappa^2$ were to be increased 
by the renormalization scheme, in order to obtain a pseudogap
in $N(\omega)$ one would likely need a large renormalization 
of the coupling $g$ and such a scheme for the present model
does not look promising to the author. However, it is important
to note that the model studied here, although similar in some
respects, is actually different than the one considered in 
Ref.\cite{Moukouri}. The renormalization scheme proposed by Moukouri
et al. which works well for the Hubbard model need not necessarily
apply to other theories. 

In Ref.\cite{PM1}, we pointed out that in the case of 
quasiparticles coupled to ferromagnetic spin fluctuations, our
results are at variance with expectations based on the standard
theory of quantum critical phenomena\cite{HertzMillis}. Since
the dynamical exponent $z=3$, in $d=2$ spatial dimensions, the
effective dimension is $d+z=5$ and is greater than the upper
critical dimension $d_c=4$ above which one would expect the
first order theory to be at least qualitatively correct. But
our nonperturbative results show that at least for small 
enough $\kappa^2$, the first order theory qualitatively breaks
down. Pepin et. al\cite{Pepin} have shown that the quasiparticle
interactions are indeed relevant in the RG sense for ferromagnetic
fluctuations in $d\leq 3$, a result consistent with our findings.
For antiferromagnetic and charge fluctuations,
$d+z=4$, the marginal case, and hence the qualitative breakdown
of the first order approximation isn't necessarily
inconsistent with the standard theory. However, the scaling
relations derived in Ref.\cite{HertzMillis} rely on the
applicability of perturbation theory. If the pseudogap effects
are indeed intrinsically nonperturbative in nature, a 
conjecture that is consistent with the present work, it 
opens the possibility that the physics in the proximity 
of a quantum critical point is dominated by nonperturbative
quantum mechanical effects and therefore even richer than
anticipated in the earlier work\cite{HertzMillis}. A number
of new ideas in this field have recently been 
proposed\cite{ColemanPepin,Si} and a discussion of some 
fundamental problems associated with quantum critical 
points can be found in Ref.\cite{QCC}.

\section{Outlook}

We studied a nonperturbative formulation of the
magnetic interaction model, in which quasiparticles
are coupled to a Gaussian distributed dynamical
molecular exchange field. Far from the magnetic
boundary, the multiple magnetic fluctuation exchange
processes do not bring about qualitative changes to the
quasiparticle spectrum. But as one gets closer to
the border of long-range magnetic order, we
find, for a range of model parameters, that Migdal's
theorem doesn't apply and the quasiparticle spectrum is
qualitatively different from its Eliashberg approximation.
Moreover, we find that going one step beyond the single
spin-fluctuation exchange approximation and including
first order vertex corrections, self-consistently or not,
doesn't help to reproduce the qualitative changes seen
in the nonperturbative calculations. Near the magnetic
boundary, the simple perturbation expansion shows signs
it is not well behaved, since the second order results
differ greatly from their first order counterparts. The
self-consistent, or renormalized perturbation expansion,
which effectively consists in a reordering of the 
diagrammatic perturbation theory, is better behaved
in that the differences between first and second order 
are much less pronounced. However, even if the renormalized 
perturbation expansion converges, our results show that
it is quite likely to converge to the wrong answer, which 
could be explained if the original perturbation expansion 
is divergent.

The intuitive argument for the onset of pseudogap 
behavior\cite{PM1,MS,Vilk,Moukouri}, namely that if the
distance quasiparticles can travel during their lifetime
becomes shorter than the molecular field correlation length,
these quasiparticles effectively see long-range order, does
not explain the failure of the single spin-fluctuation exchange
approximation. As we pointed out in Ref.\cite{PM1}, one can
get in the regime where the mean-free path gets shorter
than $1/\kappa$ in the Eliashberg approximation, but fail to
observe a pseudogap in this regime. Since we have not been
able to produce a good fit to the Monte Carlo simulations
by including either first order or vertex corrections that 
are local in space and time, i.e by a renormalization of the 
coupling constant $g$ to the molecular field, we conjecture 
that the physical origin of the pseudogap state found in the 
present calculations lies in non-local vertex corrections
produced by high order spin-fluctuation exchanges. These 
vertex corrections effectively induce a quasiparticle coupling 
to the dynamical molecular field that is non-local in both 
space and time. The above conjecture raise the question of 
what essential property of the vertex function is not captured 
by its first order approximation. 
A study of the vertex function along the same lines 
as the work reported here for the single particle Green's 
function should provide further insights into this problem. 

\section{Acknowledgments}

I would like to thank P. Coleman, J.R. Cooper, P.B. Littlewood, 
G.G. Lonzarich, J. Loram, and D. Pines for discussions on this 
and related topics. We acknowledge the support of the EPSRC, 
the Newton Trust and the Royal Society.

\begin{figure}[ht]
\protect{\includegraphics*[width=\columnwidth]{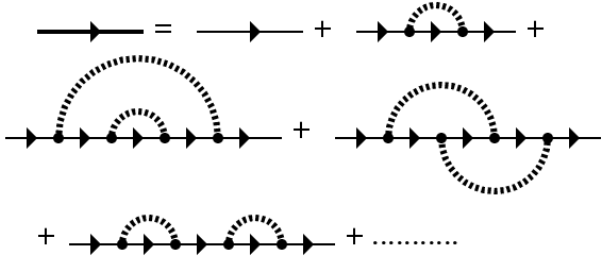}}
\protect{\caption{Diagrammatic expansion for the single particle Green's
function, Eq.(\ref{gM}) up to two spin or charge fluctuation exchanges. 
The dashed line represents the dynamical susceptibility 
$\chi({\bf q},i\nu_n)$.}}
\end{figure}

\begin{figure}[ht]
\protect{\includegraphics*[width=\columnwidth]{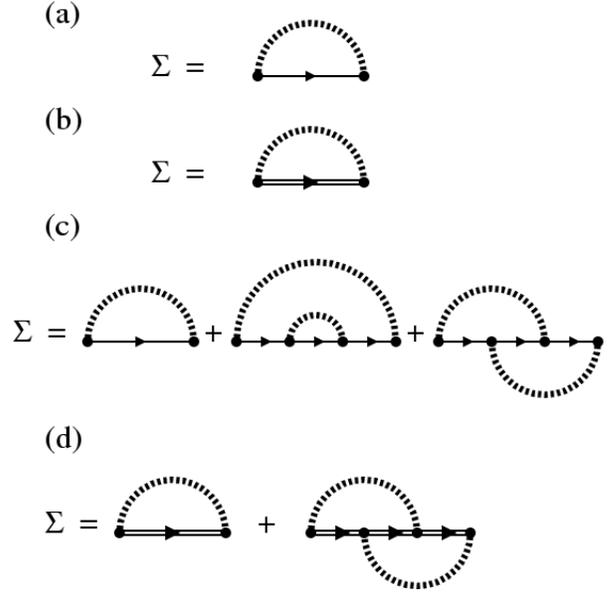}}
\protect{\caption{ The various perturbation-theoretic approximations to
the quasiparticle self-energy considered in this paper. 
(a) First order perturbation theory. (b) First order
self-consistent (Eliashberg). (c) Second order perturbation 
theory (d) Second order self-consistent. In this figure a single
line denotes the bare quasiparticle propagator $G_0$ while
a double line denotes the dressed propagator $G$ to be 
determined self-consistently.}}
\end{figure}
\clearpage

\begin{figure}[ht]
\protect{\includegraphics*[width=\columnwidth]{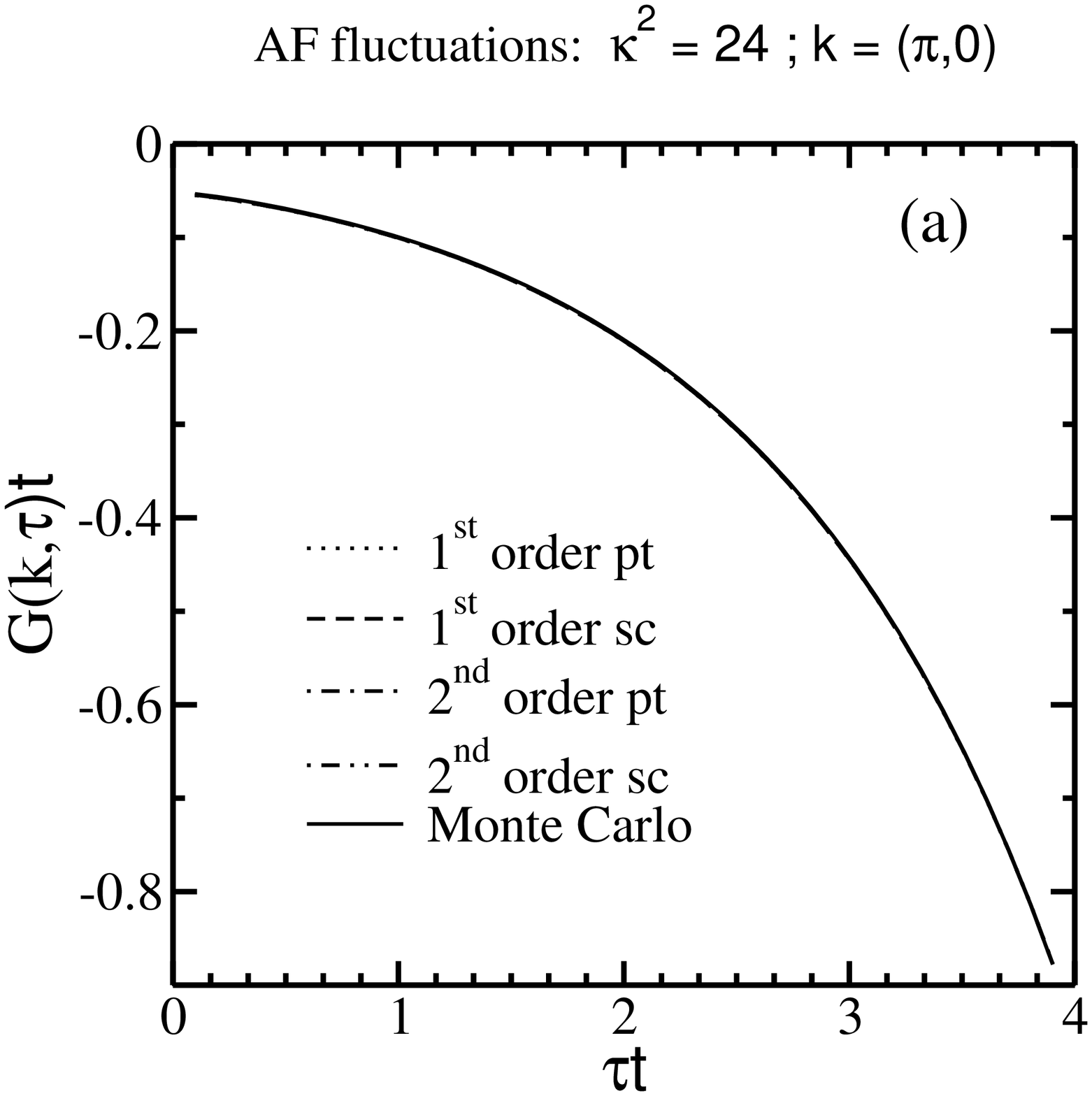}}
\end{figure}
\begin{figure}[ht]
\protect{\includegraphics*[width=\columnwidth]{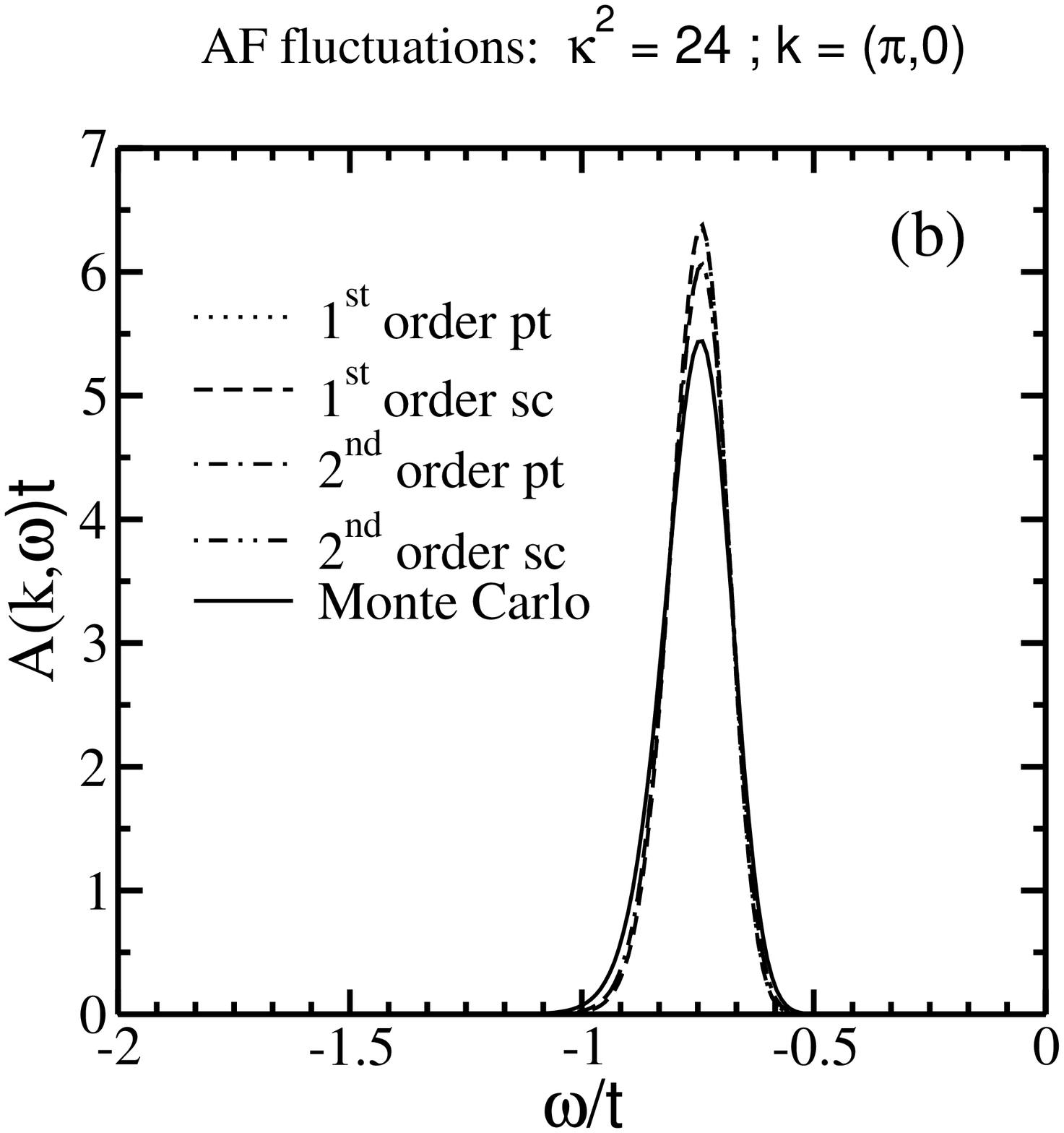}}
\end{figure}
\begin{figure}[ht]
\protect{\includegraphics*[width=\columnwidth]{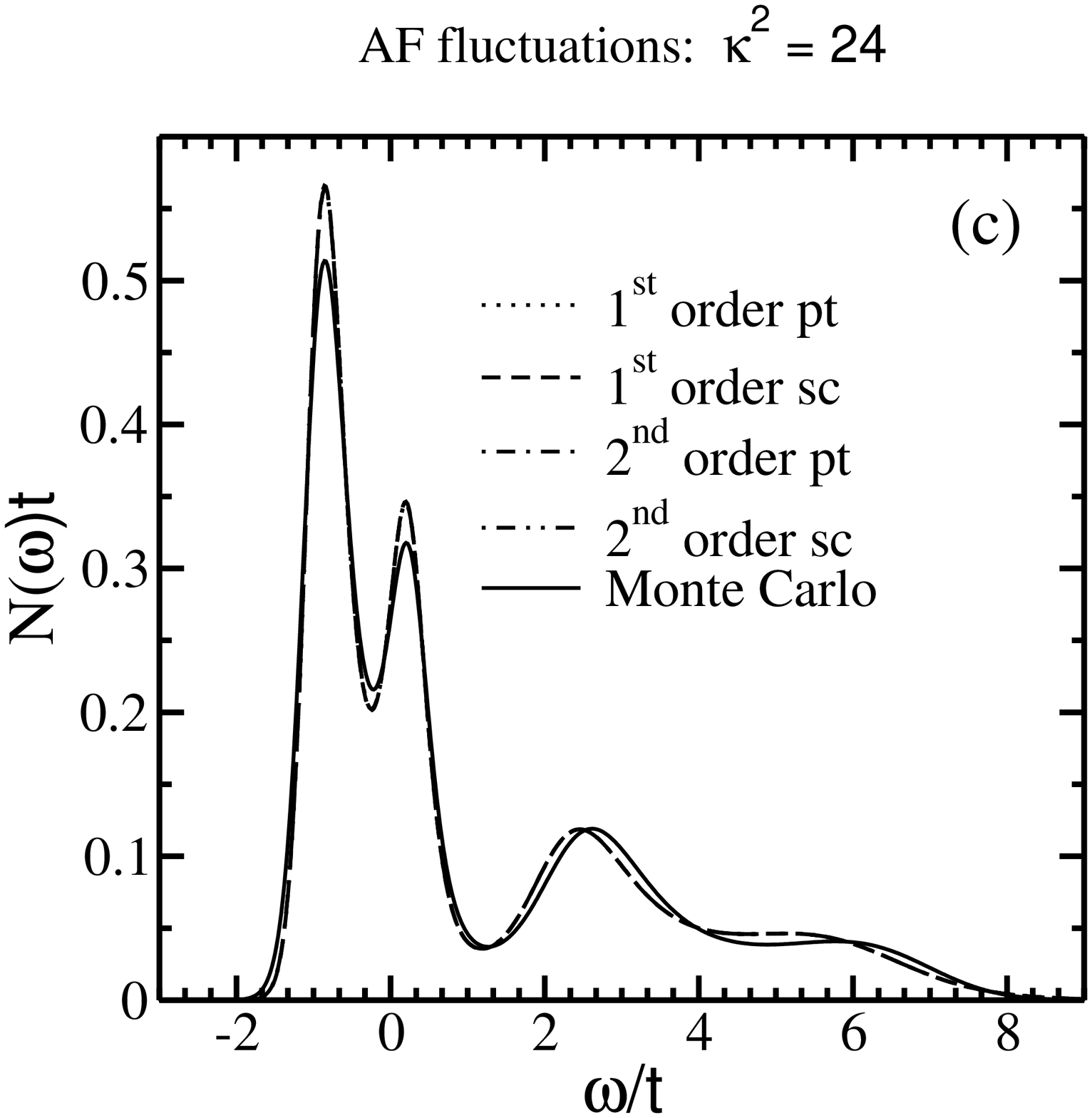}}
\protect{\caption{The various perturbation-theoretic approximations to
the quasiparticle imaginary time Green's function ${\cal G}({\bf k},\tau)$,
spectral function $A({\bf k},\omega)$ and tunneling density
of states $N(\omega)$ are compared to the results of the Monte
Carlo simulations for $\kappa^2 = 24$ in the case of quasiparticles
coupled to antiferromagnetic spin-fluctuations. $1^{st}$ order pt
corresponds to the approximation to the self-energy shown in Fig. 2a
and given by Eq.(\ref{1pt}). $2^{nd}$ order pt corresponds to 
the approximation to the self-energy shown in Fig. 2c and given 
by Eq.(\ref{2ptM}). $1^{st}$ order sc corresponds to the 
approximation to the self-energy shown in Fig. 2b and given 
by Eq.(\ref{1sc}). $2^{nd}$ order sc corresponds to the 
approximation to the self-energy shown in Fig. 2d and given by 
Eq.(\ref{2scM}). The error bars on the Monte Carlo imaginary
time Green's function are not shown for clarity. They are of
the order of 0.00002t.
(a) ${\cal G}({\bf k},\omega)$ at ${\bf k} = (\pi,0)$. 
(b) $A({\bf k},\omega)$ at ${\bf k} = (\pi,0)$. 
(c) $N(\omega)$.}}
\end{figure}
\clearpage

\begin{figure}[ht]
\protect{\includegraphics*[width=\columnwidth]{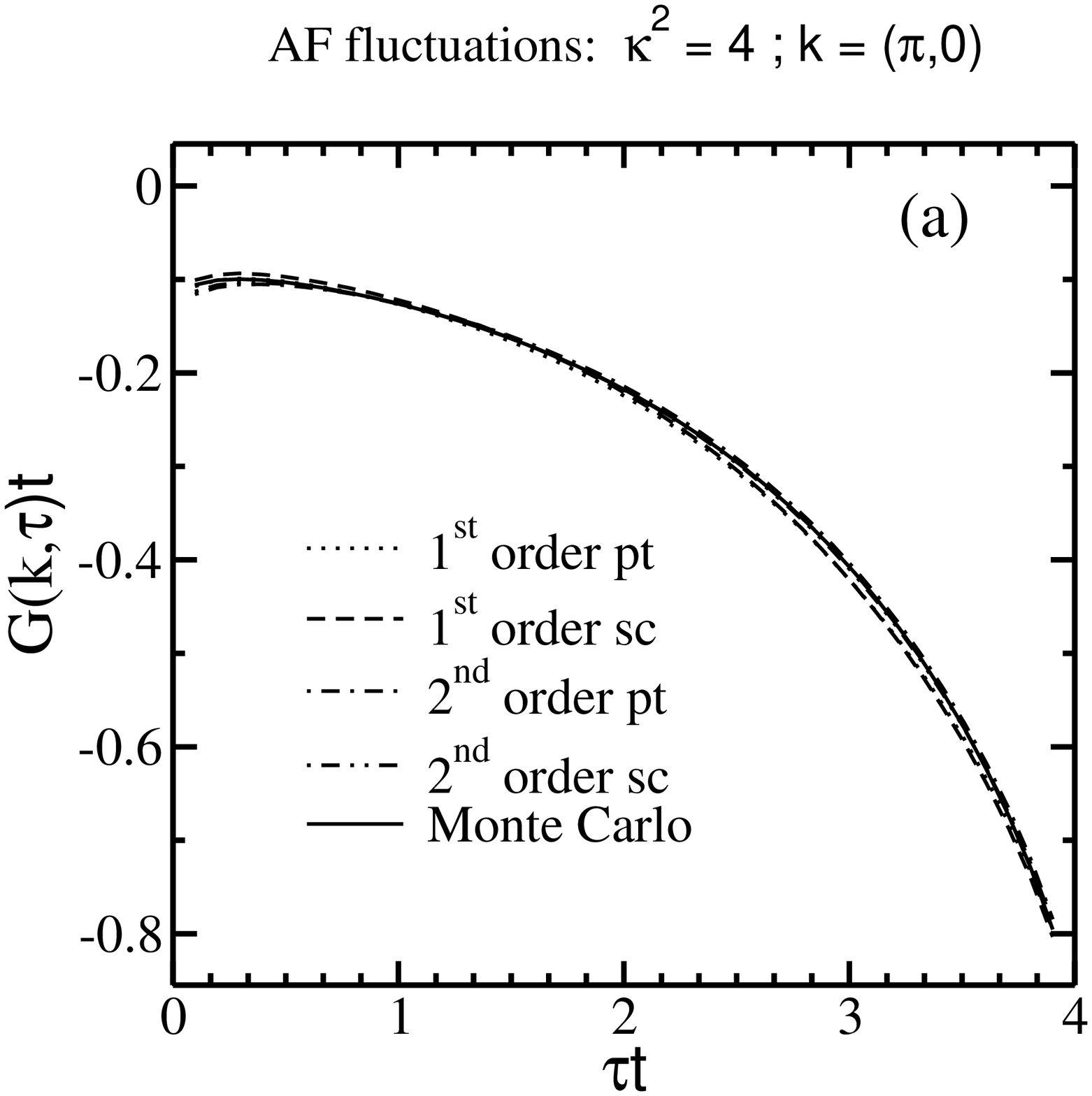}}
\end{figure}
\begin{figure}[ht]
\protect{\includegraphics*[width=\columnwidth]{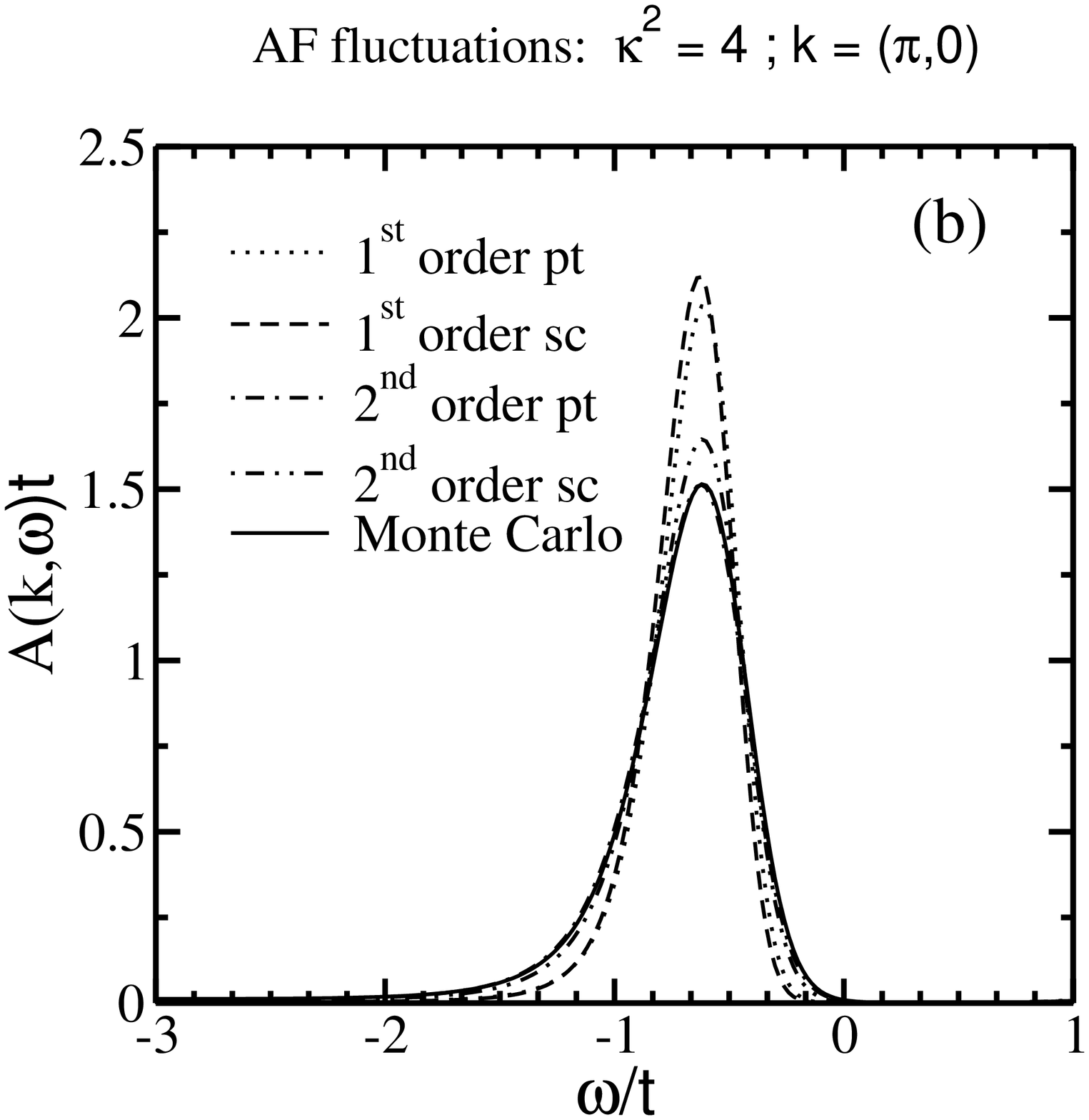}}
\end{figure}
\begin{figure}[ht]
\protect{\includegraphics*[width=\columnwidth]{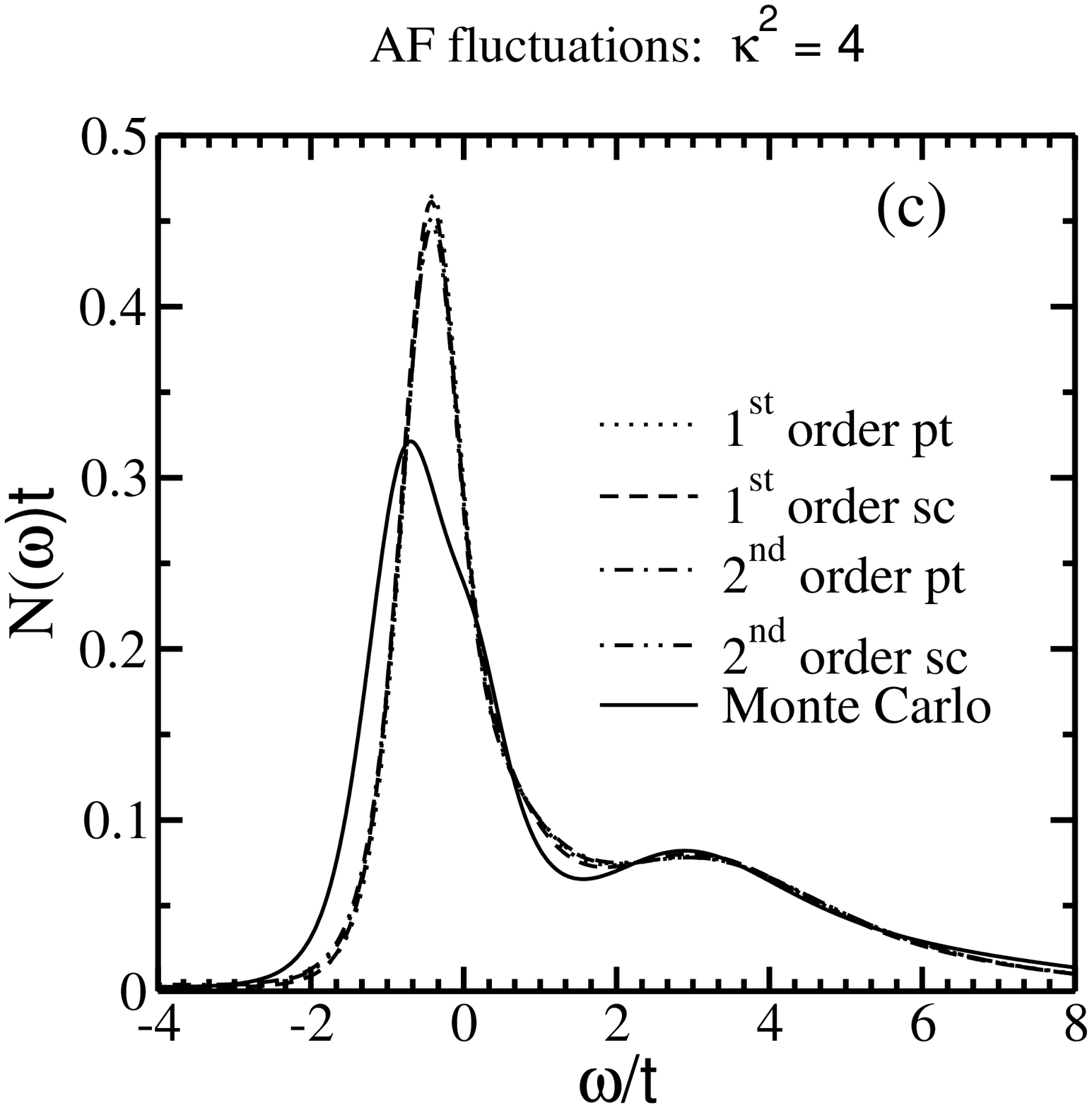}}
\protect{\caption{The various perturbation-theoretic approximations to
the quasiparticle imaginary time Green's function ${\cal G}({\bf k},\tau)$,
spectral function $A({\bf k},\omega)$ and tunneling density
of states $N(\omega)$ are compared to the results of the Monte
Carlo simulations for $\kappa^2 = 4$ in the case of quasiparticles
coupled to antiferromagnetic spin-fluctuations. $1^{st}$ order pt
corresponds to the approximation to the self-energy shown in Fig. 2a
and given by Eq.(\ref{1pt}). $2^{nd}$ order pt corresponds to 
the approximation to the self-energy shown in Fig. 2c and given 
by Eq.(\ref{2ptM}). $1^{st}$ order sc corresponds to the 
approximation to the self-energy shown in Fig. 2b and given 
by Eq.(\ref{1sc}). $2^{nd}$ order sc corresponds to the 
approximation to the self-energy shown in Fig. 2d and given by 
Eq.(\ref{2scM}). The error bars on the Monte Carlo imaginary
time Green's function are not shown for clarity. They are of
the order of 0.00015t.
(a) ${\cal G}({\bf k},\omega)$ at ${\bf k} = (\pi,0)$. 
(b) $A({\bf k},\omega)$ at ${\bf k} = (\pi,0)$. 
(c) $N(\omega)$.}}
\end{figure}
\clearpage

\begin{figure}[ht]
\protect{\includegraphics*[width=\columnwidth]{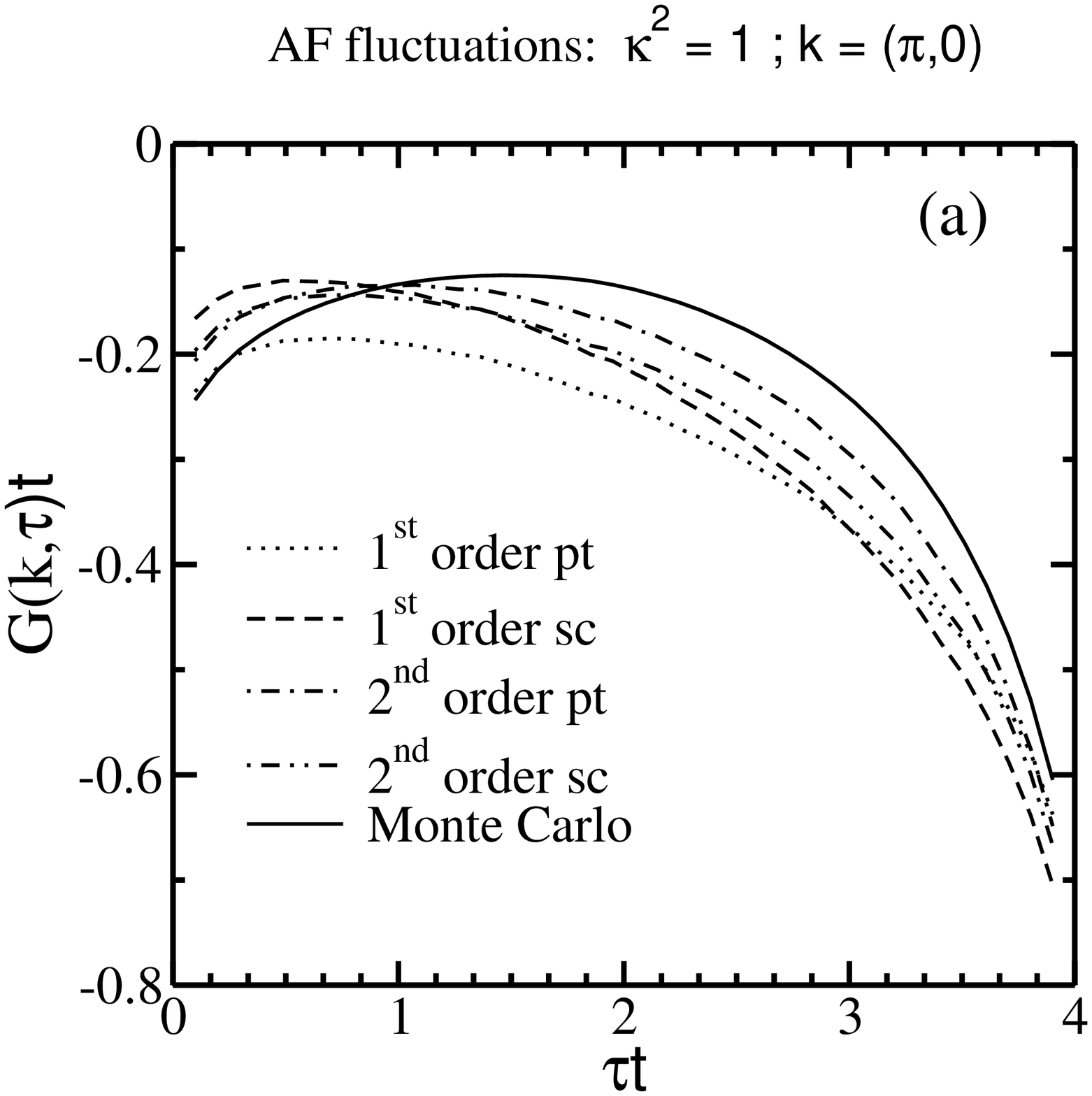}}
\end{figure}
\begin{figure}[ht]
\protect{\includegraphics*[width=\columnwidth]{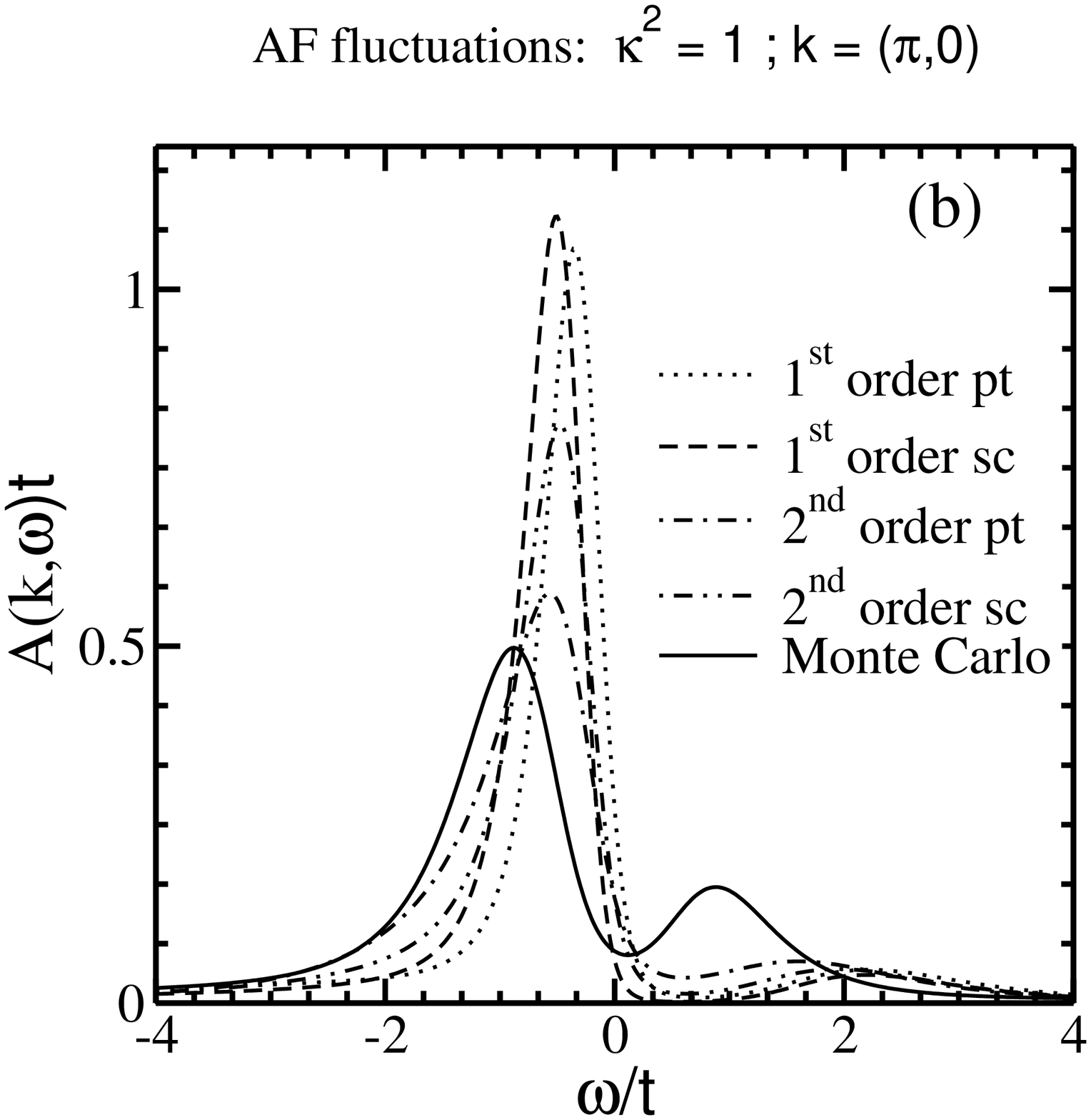}}
\end{figure}
\begin{figure}[ht]
\protect{\includegraphics*[width=\columnwidth]{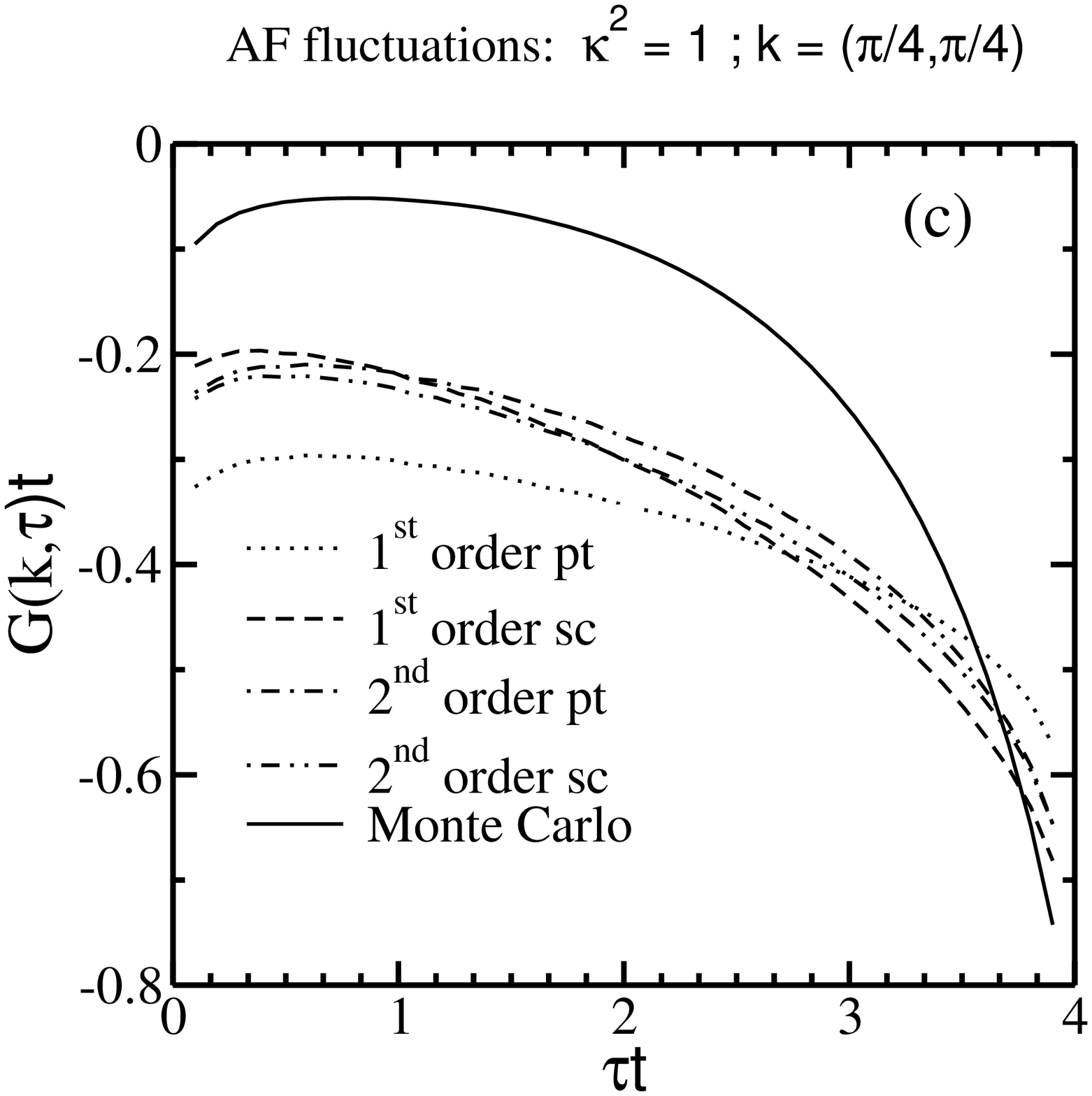}}
\end{figure}
\begin{figure}[ht]
\protect{\includegraphics*[width=\columnwidth]{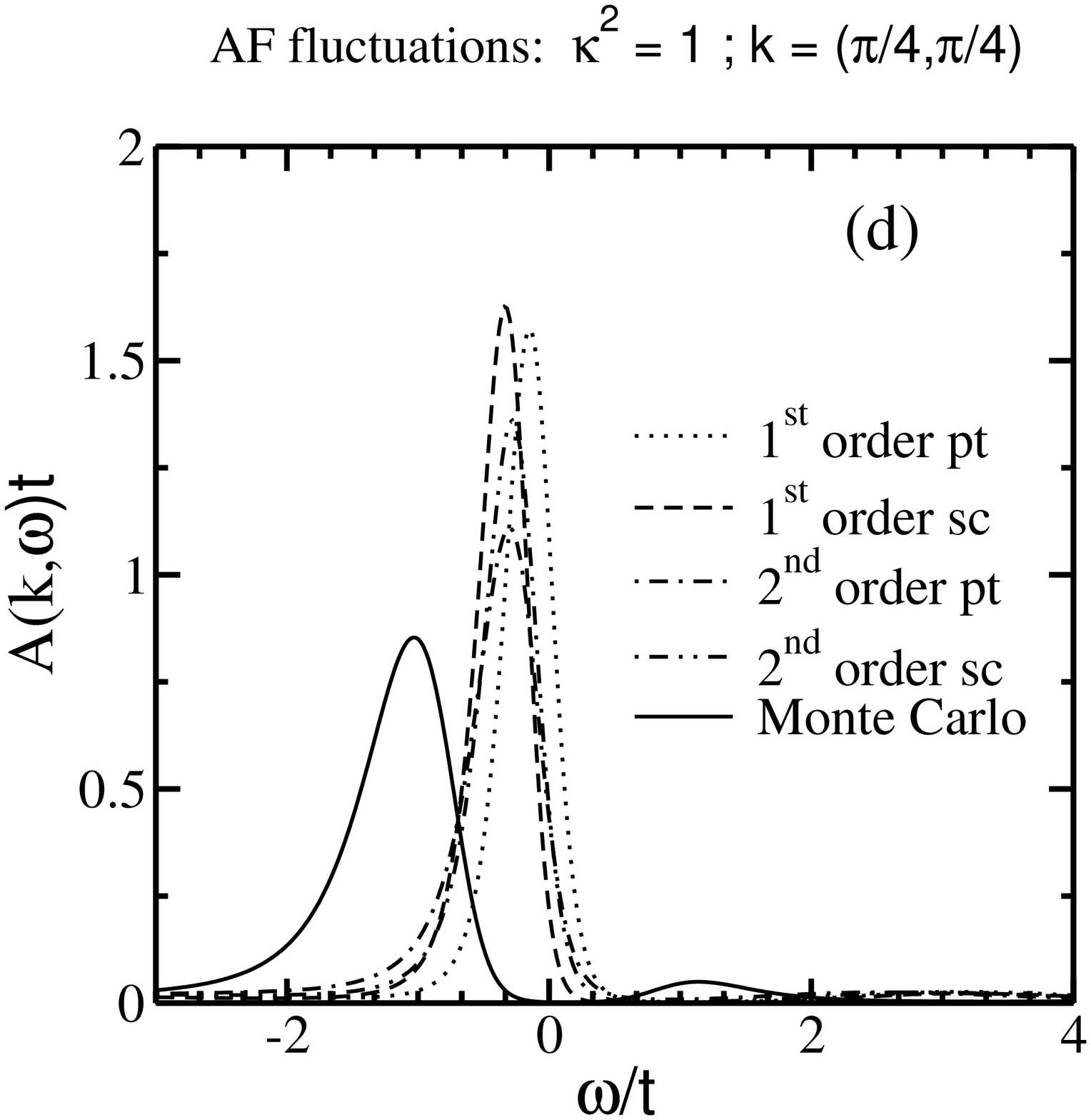}}
\end{figure}
\begin{figure}[ht]
\protect{\includegraphics*[width=\columnwidth]{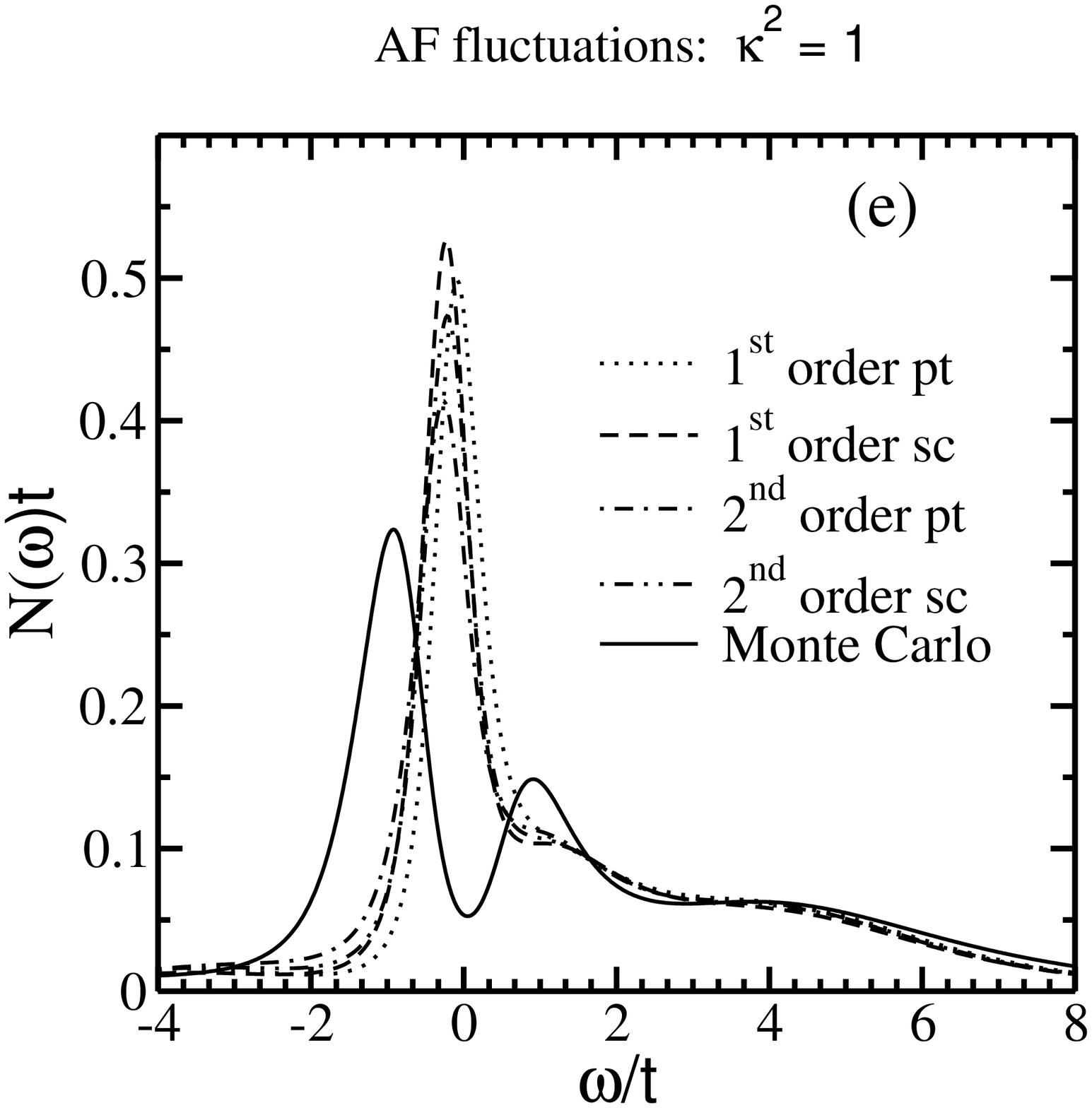}}
\protect{\caption{The various perturbation-theoretic approximations to
the quasiparticle imaginary time Green's function ${\cal G}({\bf k},\tau)$,
spectral function $A({\bf k},\omega)$ and tunneling density
of states $N(\omega)$ are compared to the results of the Monte
Carlo simulations for $\kappa^2 = 1$ in the case of quasiparticles
coupled to antiferromagnetic spin-fluctuations. $1^{st}$ order pt
corresponds to the approximation to the self-energy shown in Fig. 2a
and given by Eq.(\ref{1pt}). $2^{nd}$ order pt corresponds to 
the approximation to the self-energy shown in Fig. 2c and given 
by Eq.(\ref{2ptM}). $1^{st}$ order sc corresponds to the 
approximation to the self-energy shown in Fig. 2b and given 
by Eq.(\ref{1sc}). $2^{nd}$ order sc corresponds to the 
approximation to the self-energy shown in Fig. 2d and given by 
Eq.(\ref{2scM}). The error bars on the Monte Carlo imaginary
time Green's function are not shown for clarity. They are of
the order of 0.001t.
(a) ${\cal G}({\bf k},\omega)$ at ${\bf k} = (\pi,0)$. 
(b) $A({\bf k},\omega)$ at ${\bf k} = (\pi,0)$. 
(c) ${\cal G}({\bf k},\omega)$ at ${\bf k} = (\pi/4,\pi/4)$. 
(d) $A({\bf k},\omega)$ at ${\bf k} = (\pi/4,\pi/4)$. 
(e) $N(\omega)$.}}
\end{figure}
\clearpage

\begin{figure}[ht]
\protect{\includegraphics*[width=\columnwidth]{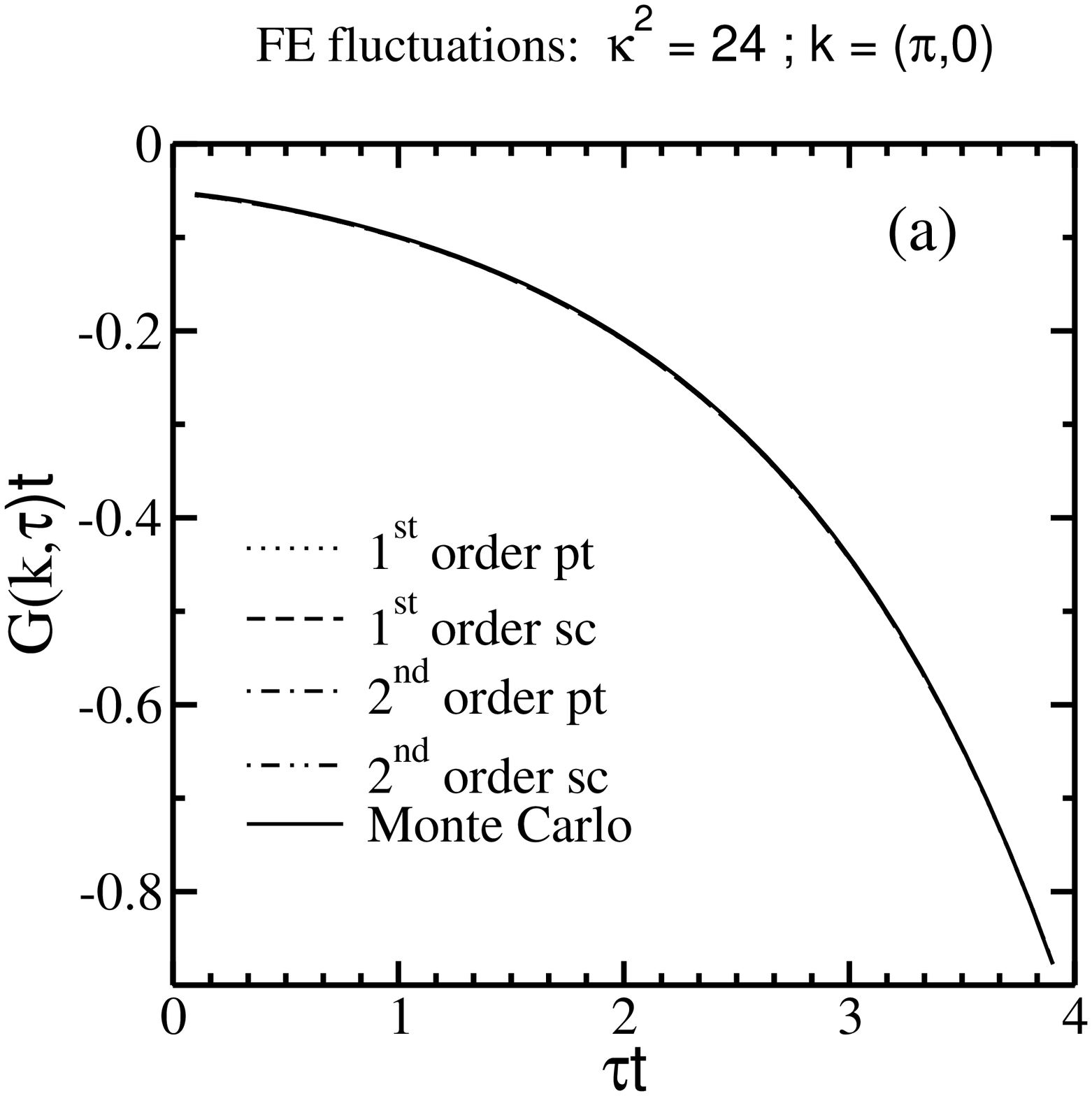}}
\end{figure}
\begin{figure}[ht]
\protect{\includegraphics*[width=\columnwidth]{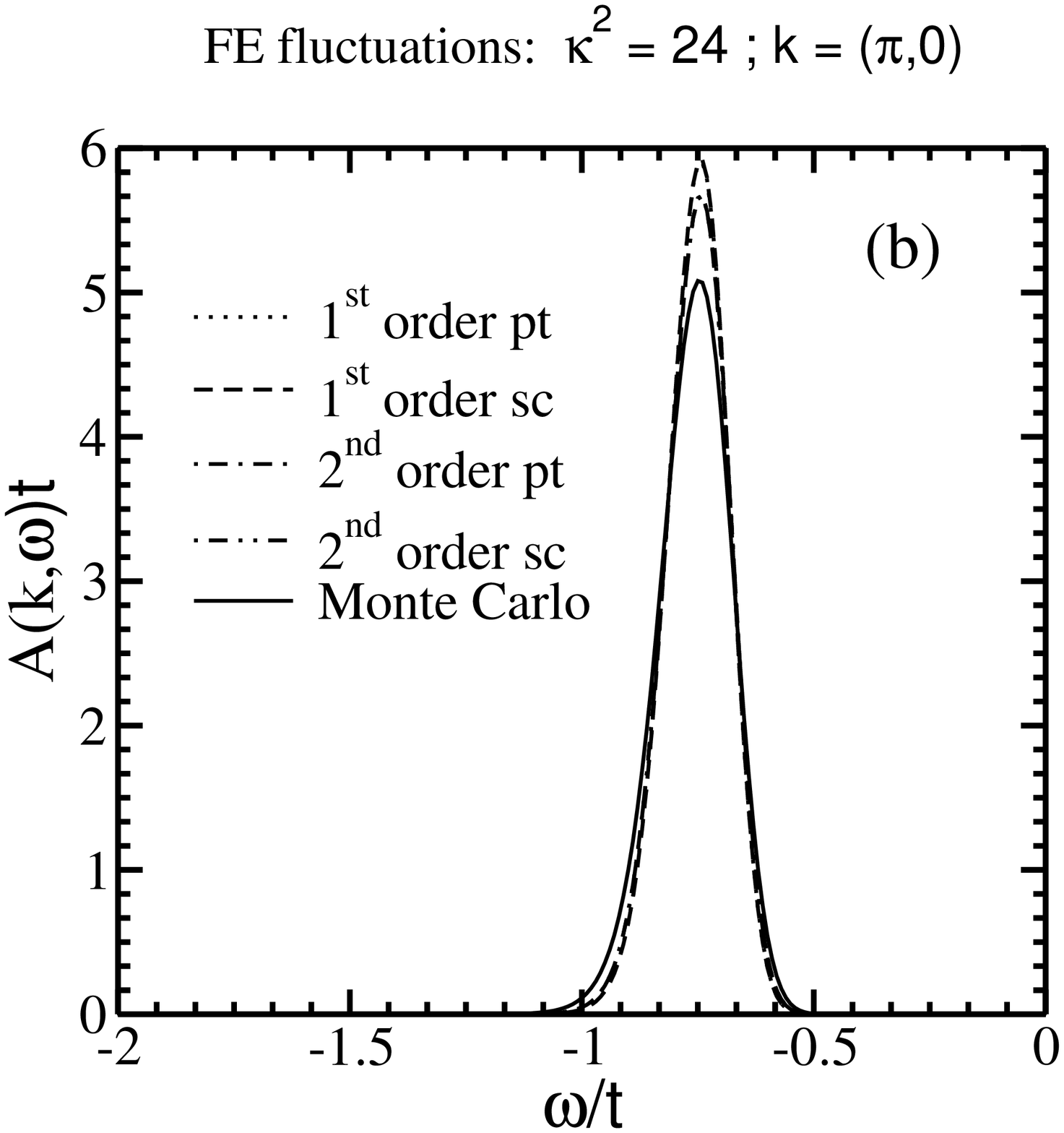}}
\end{figure}
\begin{figure}[ht]
\protect{\includegraphics*[width=\columnwidth]{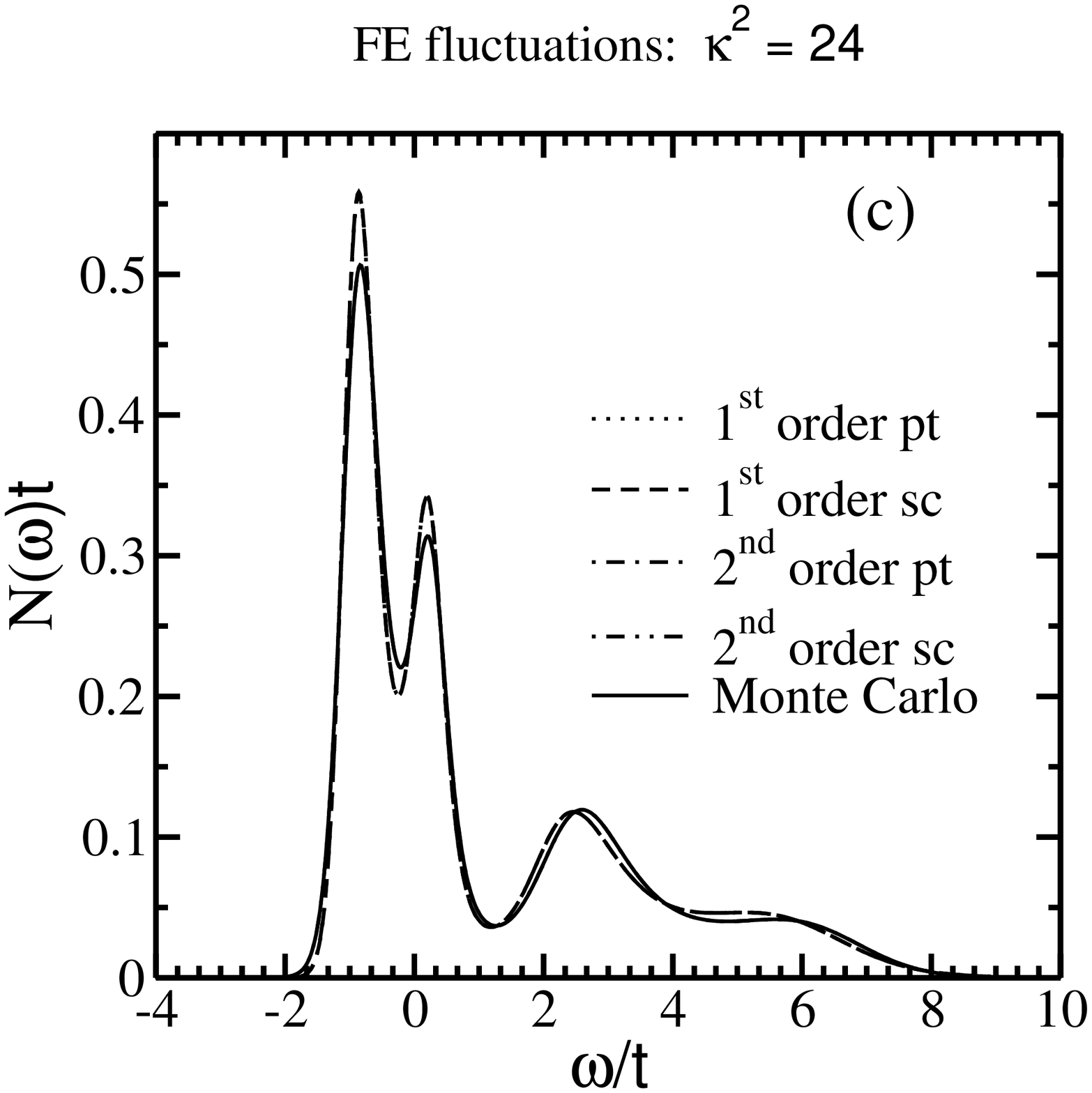}}
\protect{\caption{The various perturbation-theoretic approximations to
the quasiparticle imaginary time Green's function ${\cal G}({\bf k},\tau)$,
spectral function $A({\bf k},\omega)$ and tunneling density
of states $N(\omega)$ are compared to the results of the Monte
Carlo simulations for $\kappa^2 = 24$ in the case of quasiparticles
coupled to ferromagnetic spin-fluctuations. $1^{st}$ order pt
corresponds to the approximation to the self-energy shown in Fig. 2a
and given by Eq.(\ref{1pt}). $2^{nd}$ order pt corresponds to 
the approximation to the self-energy shown in Fig. 2c and given 
by Eq.(\ref{2ptM}). $1^{st}$ order sc corresponds to the 
approximation to the self-energy shown in Fig. 2b and given 
by Eq.(\ref{1sc}). $2^{nd}$ order sc corresponds to the 
approximation to the self-energy shown in Fig. 2d and given by 
Eq.(\ref{2scM}). The error bars on the Monte Carlo imaginary
time Green's function are not shown for clarity. They are of
the order of 0.00002t.
(a) ${\cal G}({\bf k},\omega)$ at ${\bf k} = (\pi,0)$. 
(b) $A({\bf k},\omega)$ at ${\bf k} = (\pi,0)$. 
(c) $N(\omega)$.}}
\end{figure}
\clearpage

\begin{figure}[ht]
\protect{\includegraphics*[width=\columnwidth]{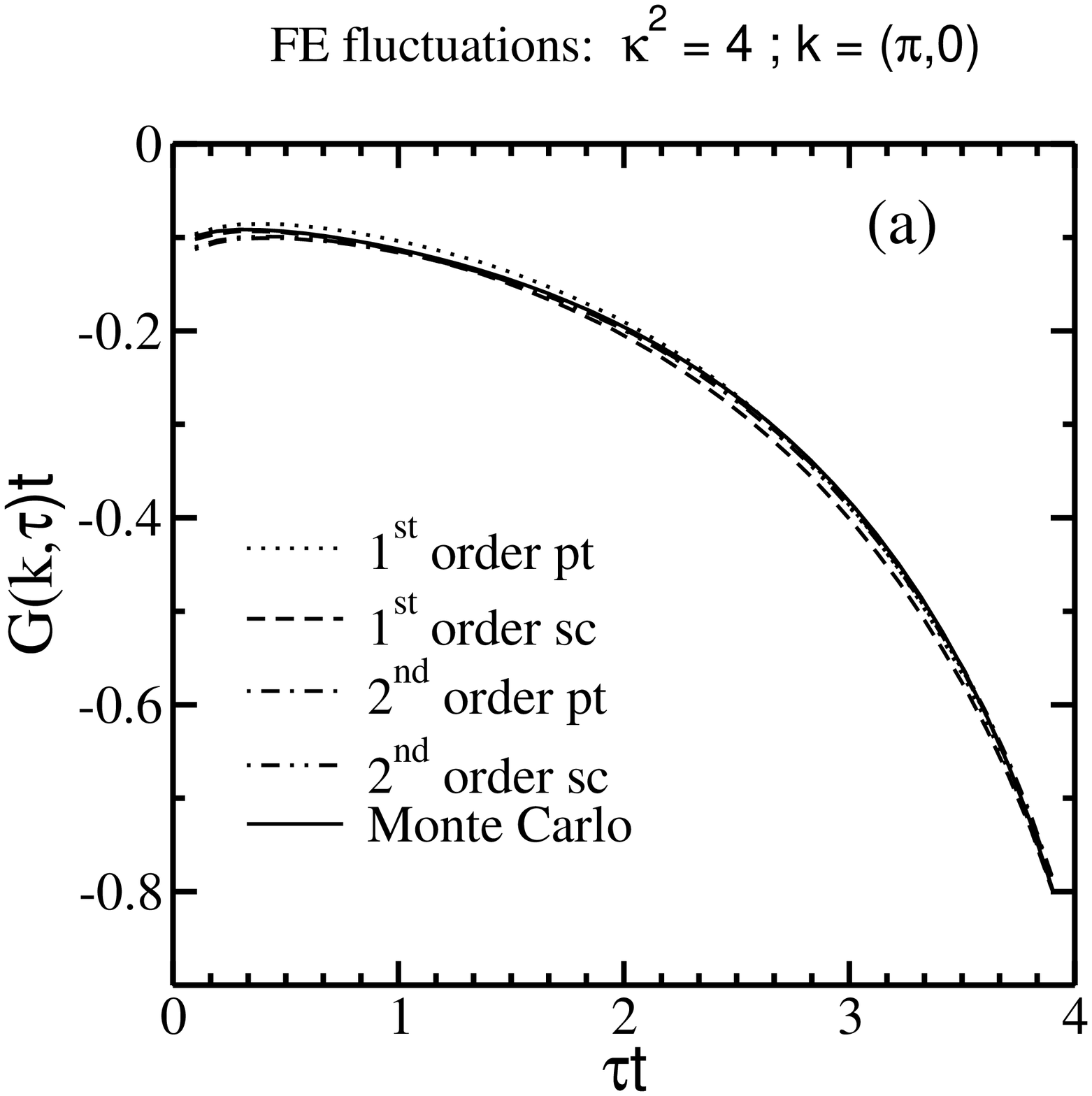}}
\end{figure}
\begin{figure}[ht]
\protect{\includegraphics*[width=\columnwidth]{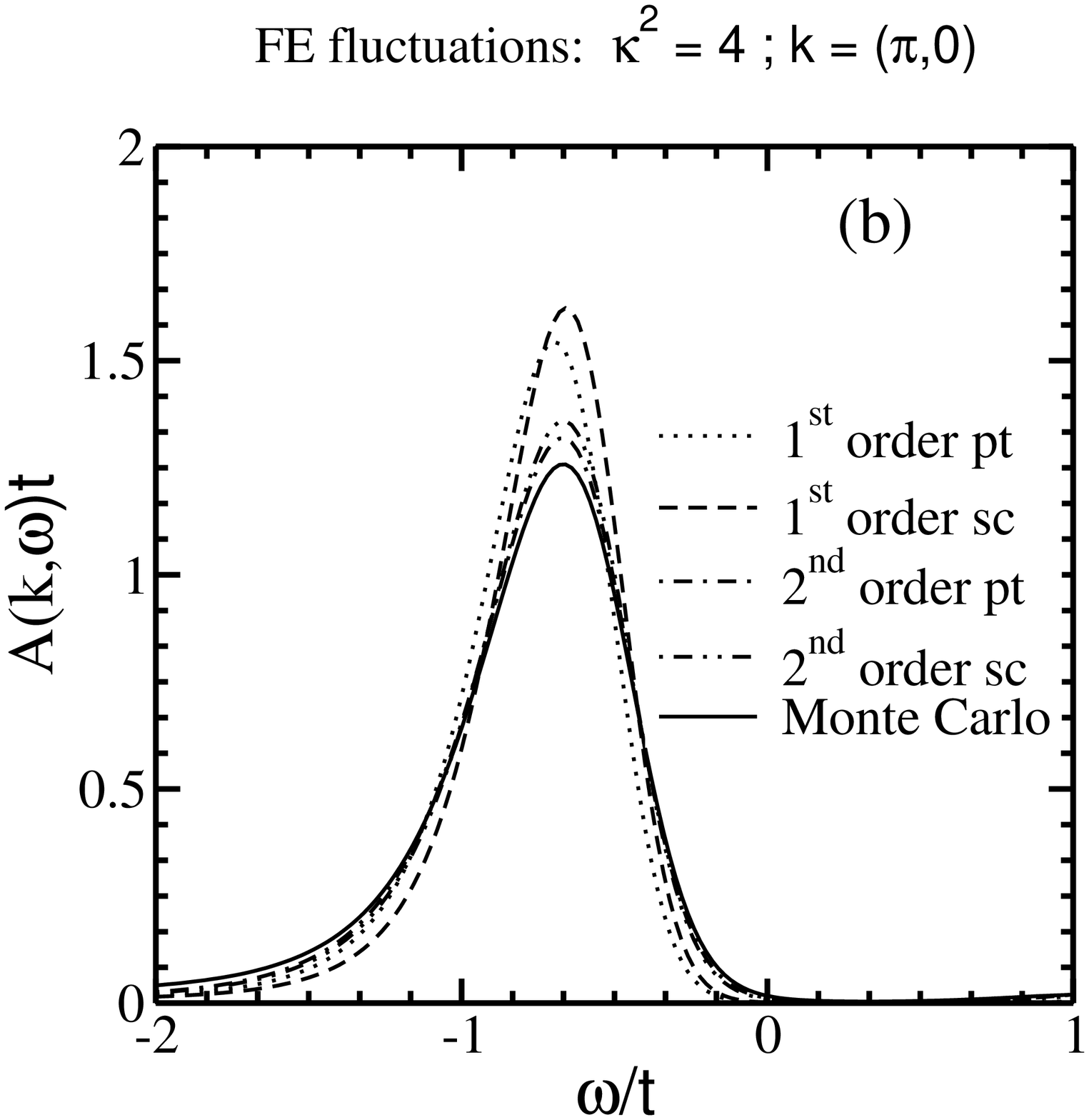}}
\end{figure}
\begin{figure}[ht]
\protect{\includegraphics*[width=\columnwidth]{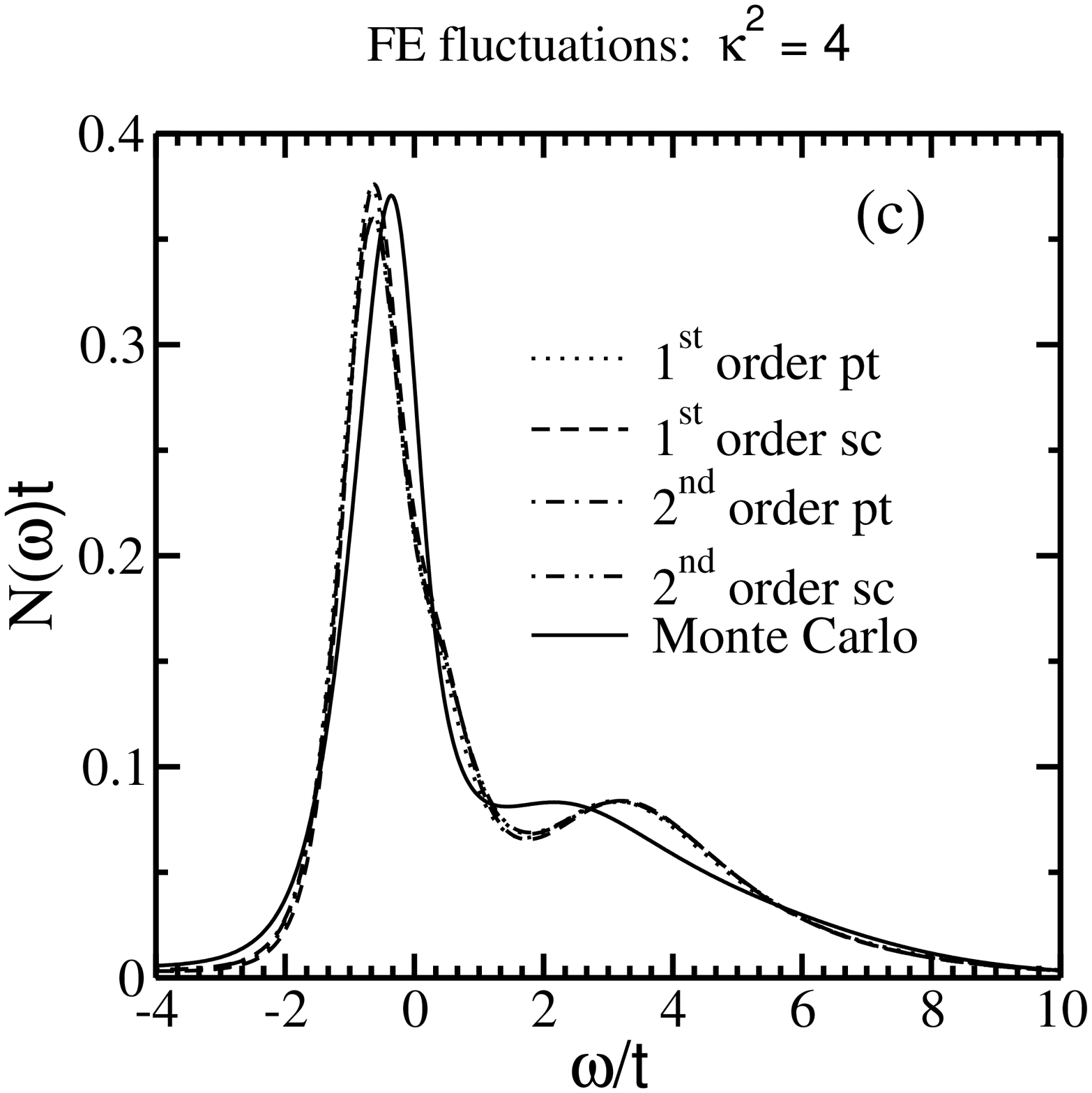}}
\protect{\caption{The various perturbation-theoretic approximations to
the quasiparticle imaginary time Green's function ${\cal G}({\bf k},\tau)$,
spectral function $A({\bf k},\omega)$ and tunneling density
of states $N(\omega)$ are compared to the results of the Monte
Carlo simulations for $\kappa^2 = 4$ in the case of quasiparticles
coupled to ferromagnetic spin-fluctuations. $1^{st}$ order pt
corresponds to the approximation to the self-energy shown in Fig. 2a
and given by Eq.(\ref{1pt}). $2^{nd}$ order pt corresponds to 
the approximation to the self-energy shown in Fig. 2c and given 
by Eq.(\ref{2ptM}). $1^{st}$ order sc corresponds to the 
approximation to the self-energy shown in Fig. 2b and given 
by Eq.(\ref{1sc}). $2^{nd}$ order sc corresponds to the 
approximation to the self-energy shown in Fig. 2d and given by 
Eq.(\ref{2scM}). The error bars on the Monte Carlo imaginary
time Green's function are not shown for clarity. They are of
the order of 0.0002t.
(a) ${\cal G}({\bf k},\omega)$ at ${\bf k} = (\pi,0)$. 
(b) $A({\bf k},\omega)$ at ${\bf k} = (\pi,0)$. 
(c) $N(\omega)$.}}
\end{figure}
\clearpage

\begin{figure}[ht]
\protect{\includegraphics*[width=\columnwidth]{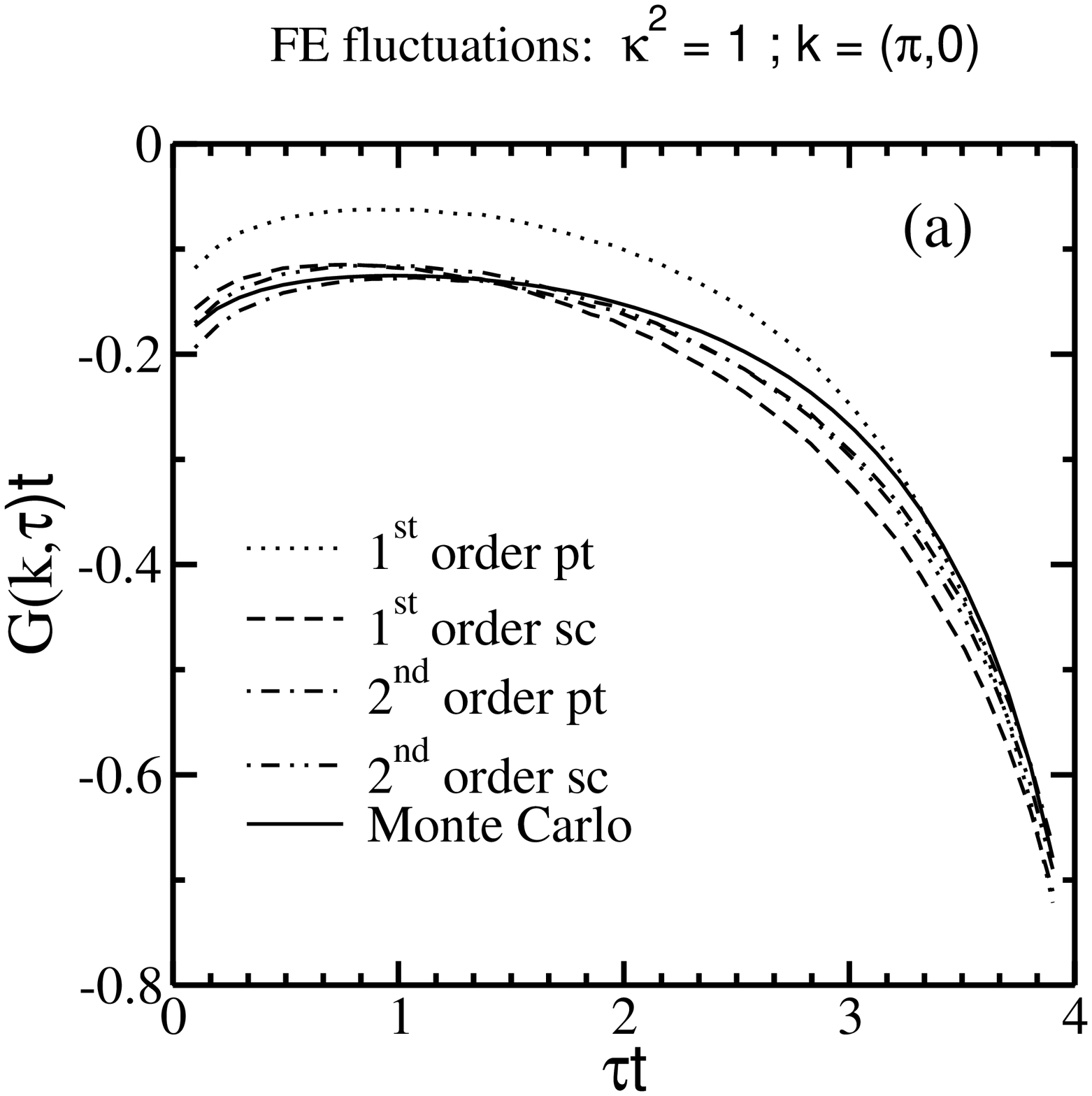}}
\end{figure}
\begin{figure}[ht]
\protect{\includegraphics*[width=\columnwidth]{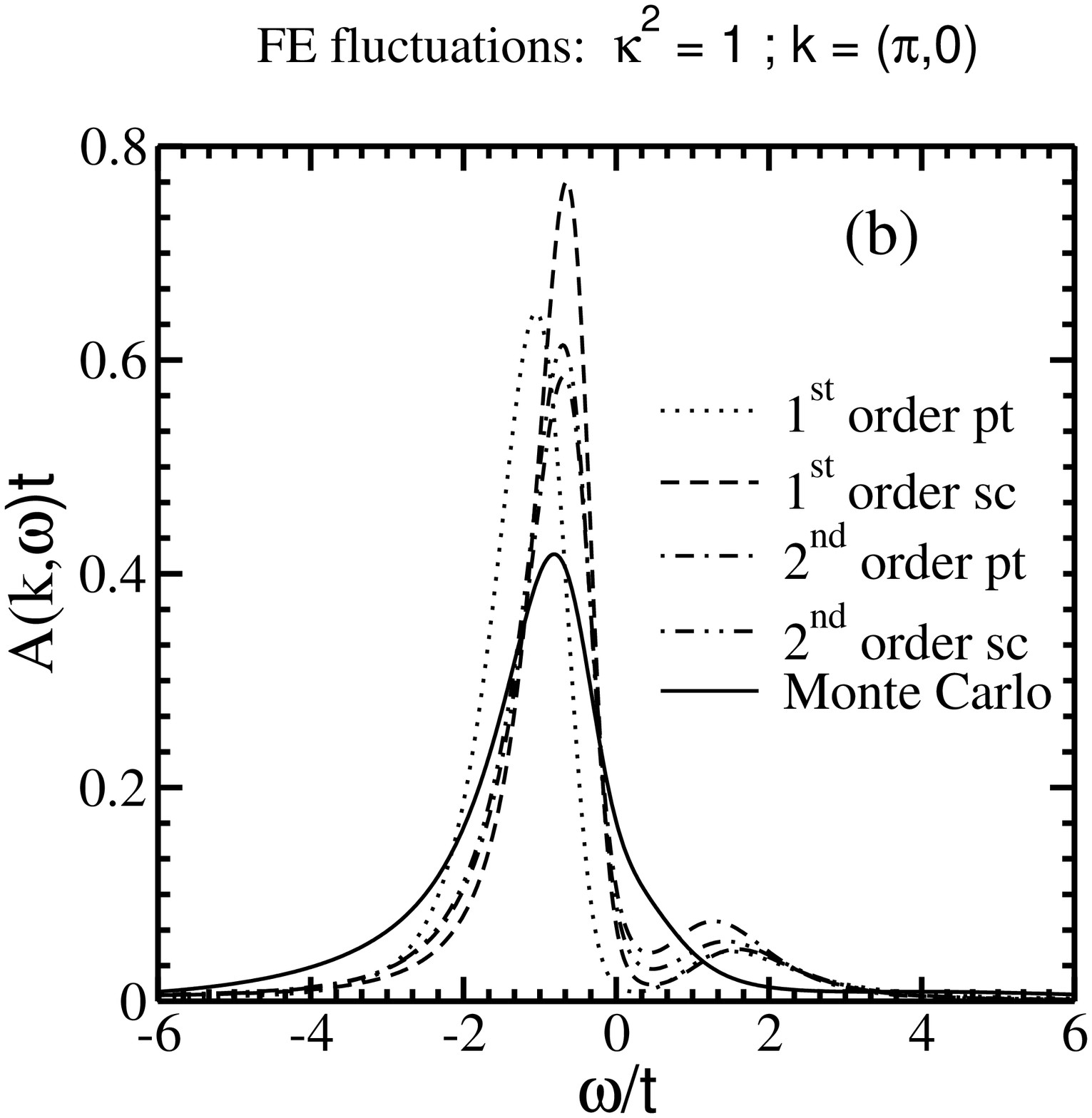}}
\end{figure}
\begin{figure}[ht]
\protect{\includegraphics*[width=\columnwidth]{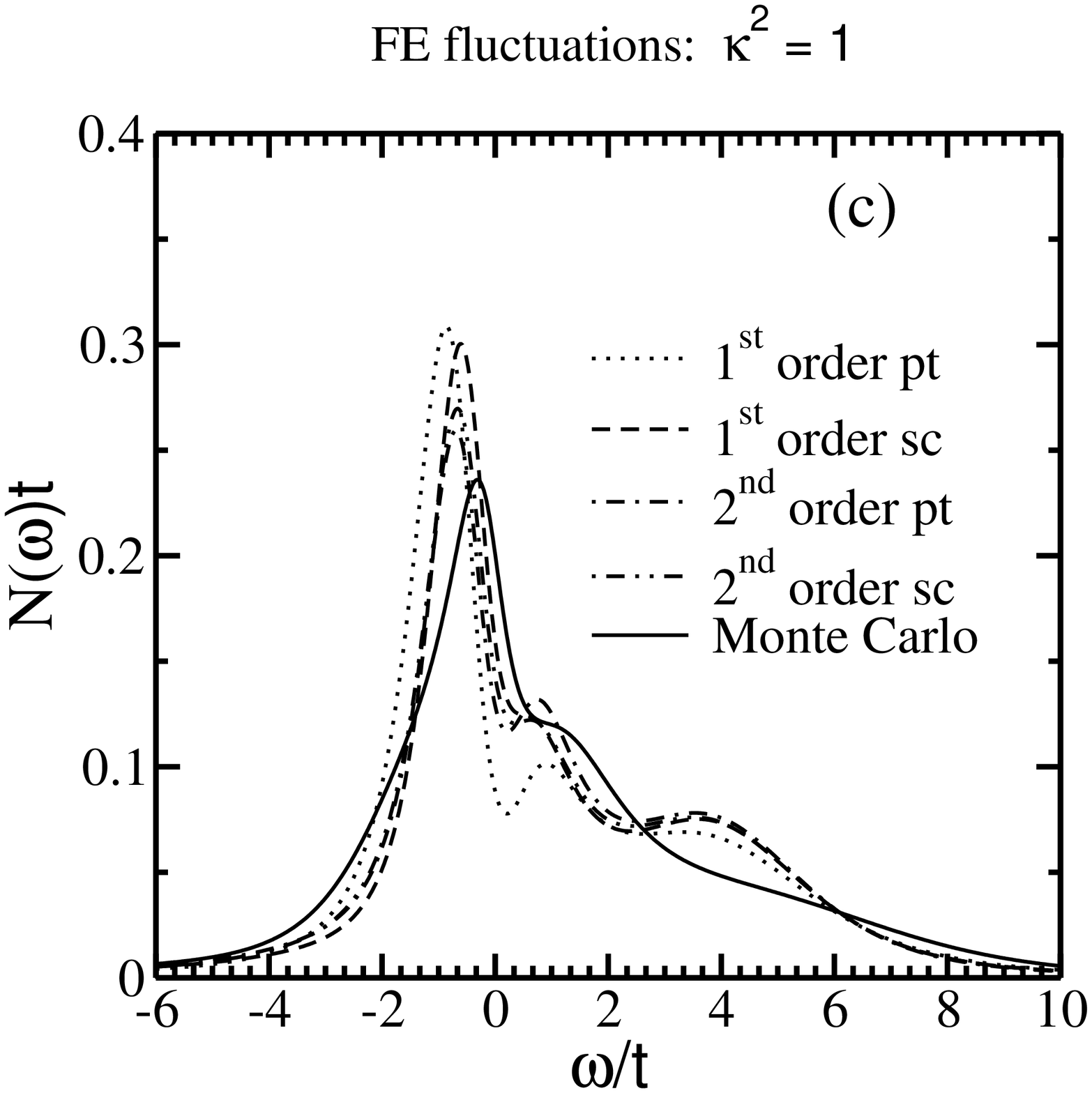}}
\protect{\caption{The various perturbation-theoretic approximations to
the quasiparticle imaginary time Green's function ${\cal G}({\bf k},\tau)$,
spectral function $A({\bf k},\omega)$ and tunneling density
of states $N(\omega)$ are compared to the results of the Monte
Carlo simulations for $\kappa^2 = 1$ in the case of quasiparticles
coupled to ferromagnetic spin-fluctuations. $1^{st}$ order pt
corresponds to the approximation to the self-energy shown in Fig. 2a
and given by Eq.(\ref{1pt}). $2^{nd}$ order pt corresponds to 
the approximation to the self-energy shown in Fig. 2c and given 
by Eq.(\ref{2ptM}). $1^{st}$ order sc corresponds to the 
approximation to the self-energy shown in Fig. 2b and given 
by Eq.(\ref{1sc}). $2^{nd}$ order sc corresponds to the 
approximation to the self-energy shown in Fig. 2d and given by 
Eq.(\ref{2scM}). The error bars on the Monte Carlo imaginary
time Green's function are not shown for clarity. They are of
the order of 0.0008t.
(a) ${\cal G}({\bf k},\omega)$ at ${\bf k} = (\pi,0)$. 
(b) $A({\bf k},\omega)$ at ${\bf k} = (\pi,0)$. 
(c) $N(\omega)$.}}
\end{figure}
\clearpage

\begin{figure}[ht]
\protect{\includegraphics*[width=\columnwidth]{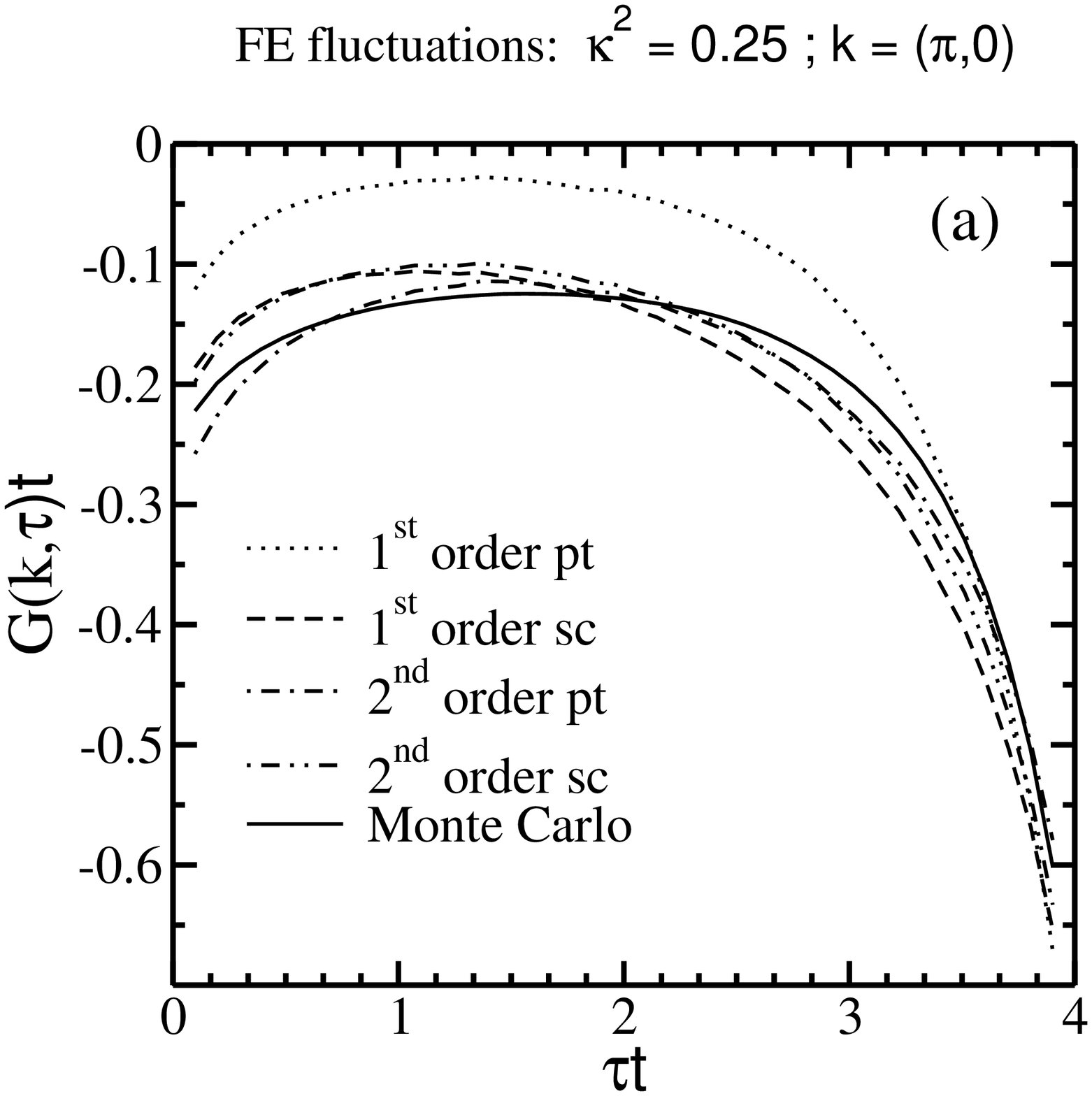}}
\end{figure}
\begin{figure}[ht]
\protect{\includegraphics*[width=\columnwidth]{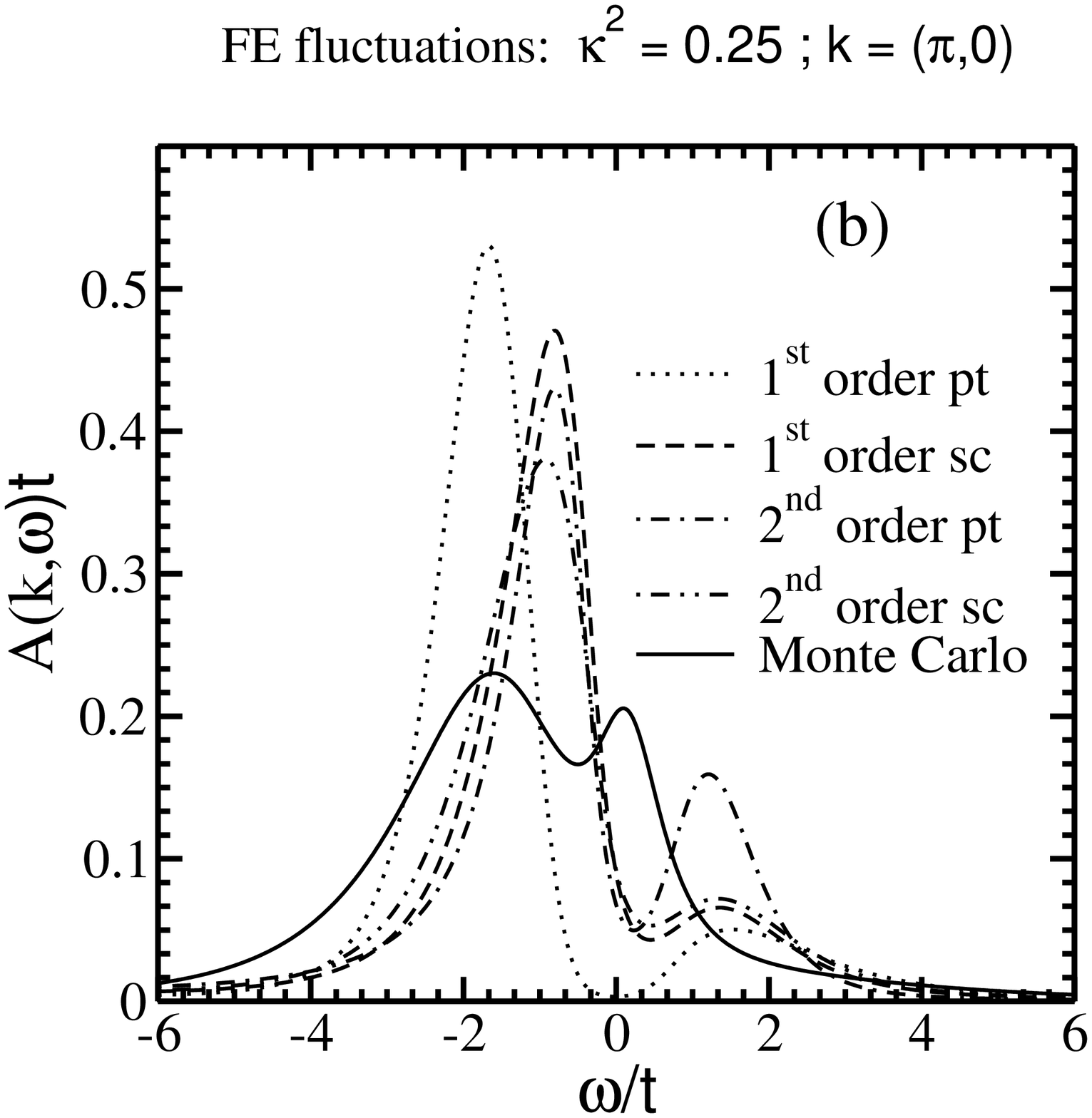}}
\end{figure}
\begin{figure}[ht]
\protect{\includegraphics*[width=\columnwidth]{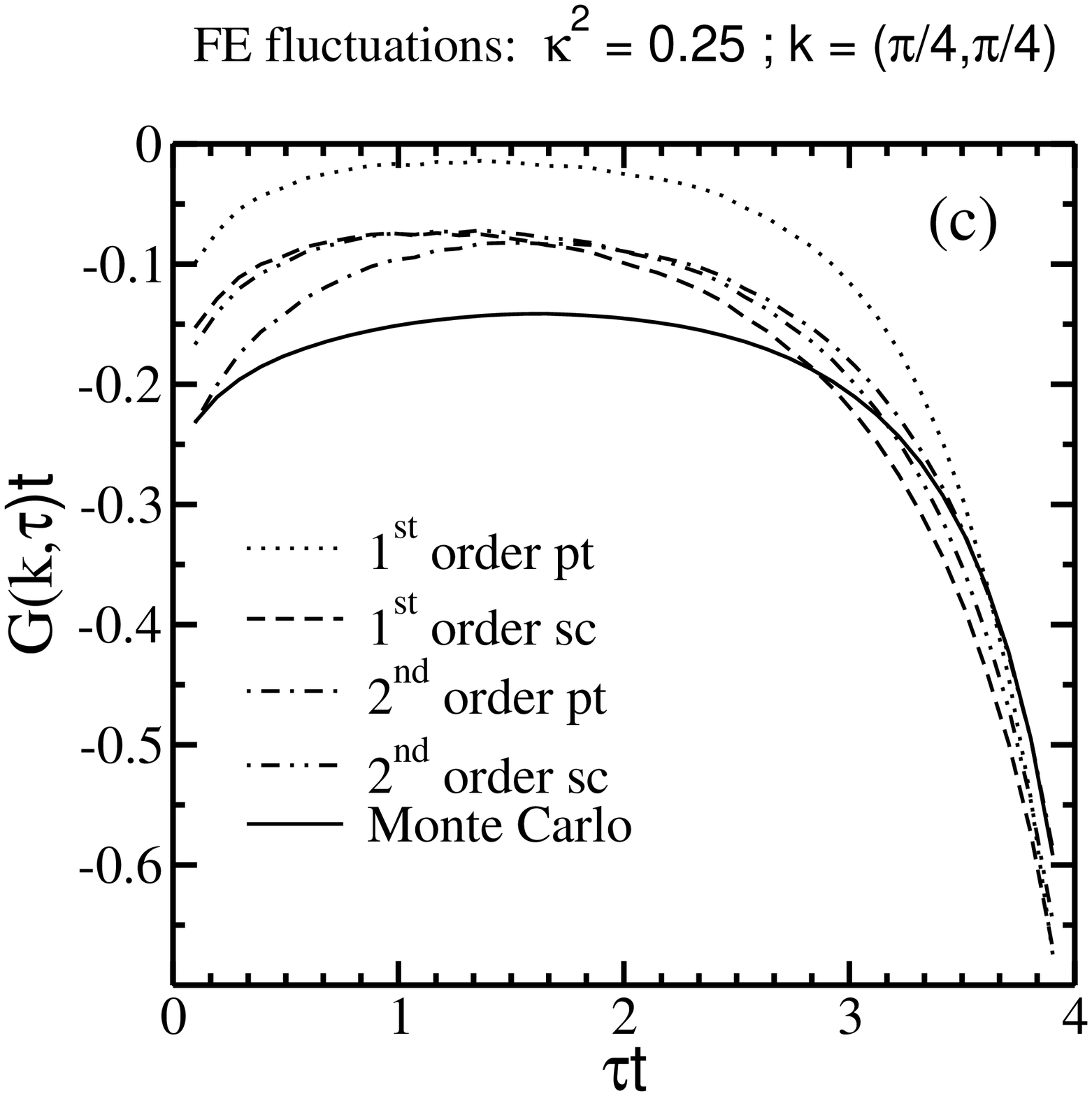}}
\end{figure}
\begin{figure}[ht]
\protect{\includegraphics*[width=\columnwidth]{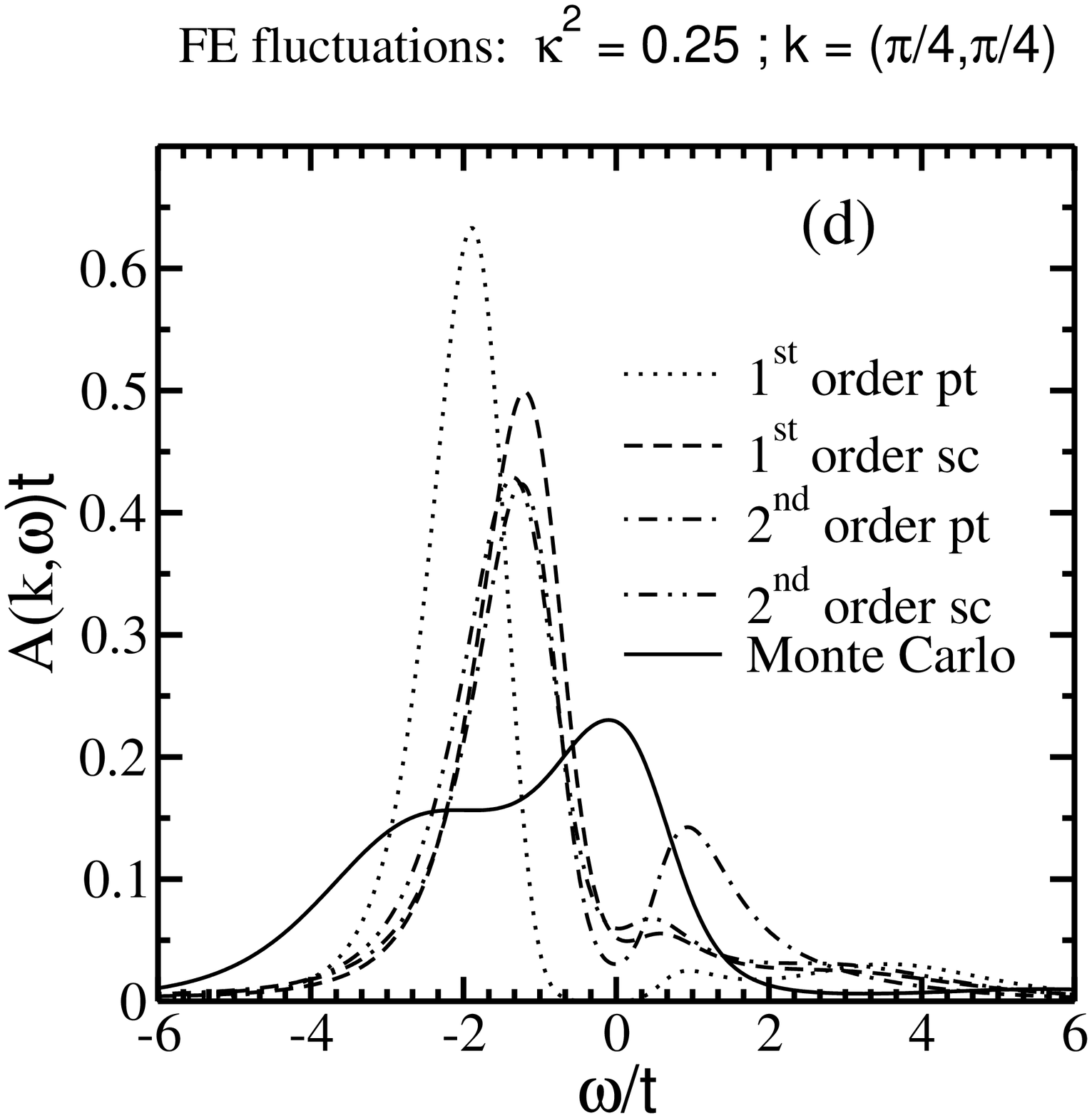}}
\end{figure}
\begin{figure}[ht]
\protect{\includegraphics*[width=\columnwidth]{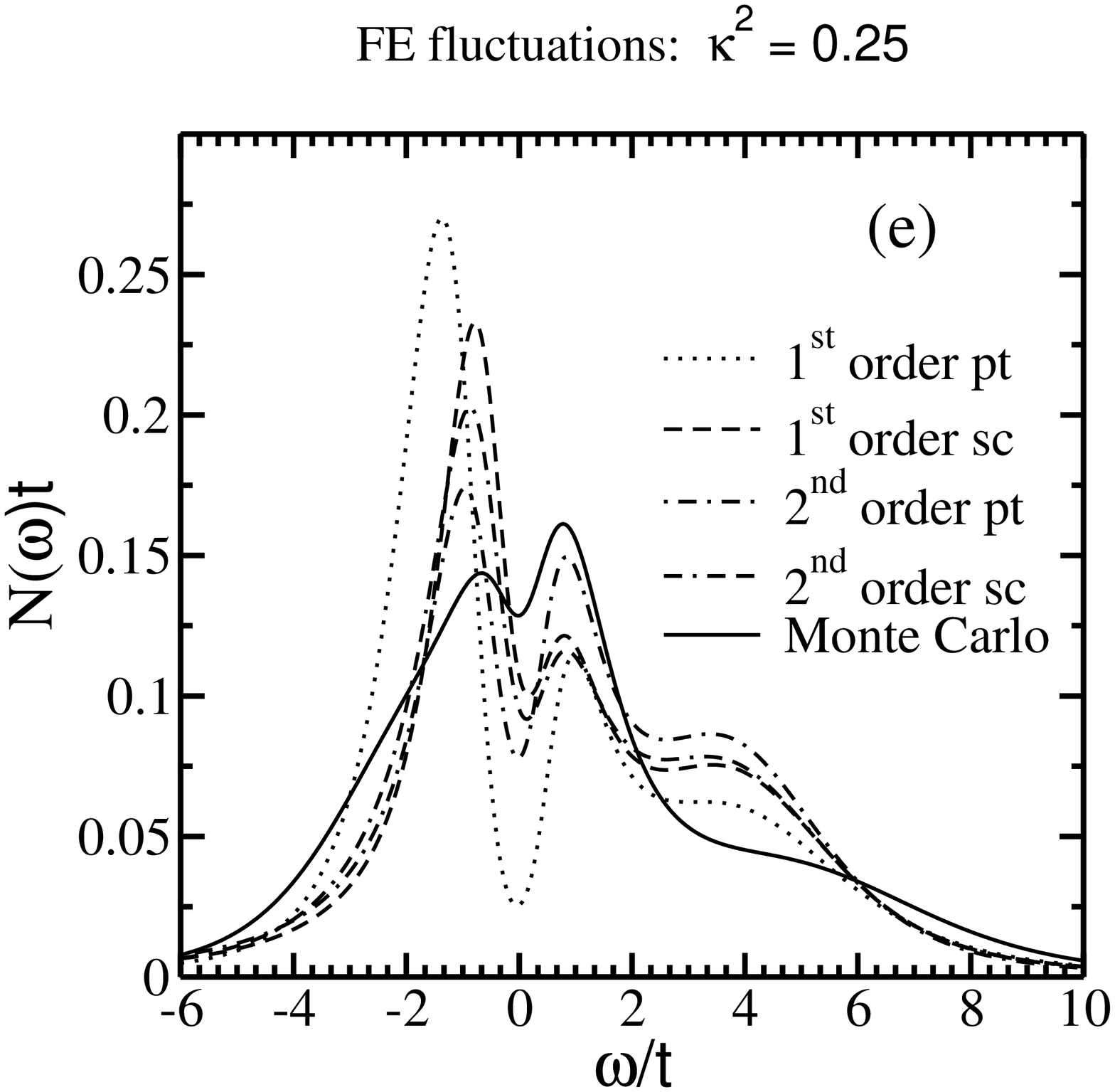}}
\protect{\caption{The various perturbation-theoretic approximations to
the quasiparticle imaginary time Green's function ${\cal G}({\bf k},\tau)$,
spectral function $A({\bf k},\omega)$ and tunneling density
of states $N(\omega)$ are compared to the results of the Monte
Carlo simulations for $\kappa^2 = 0.25$ in the case of quasiparticles
coupled to ferromagnetic spin-fluctuations. $1^{st}$ order pt
corresponds to the approximation to the self-energy shown in Fig. 2a
and given by Eq.(\ref{1pt}). $2^{nd}$ order pt corresponds to 
the approximation to the self-energy shown in Fig. 2c and given 
by Eq.(\ref{2ptM}). $1^{st}$ order sc corresponds to the 
approximation to the self-energy shown in Fig. 2b and given 
by Eq.(\ref{1sc}). $2^{nd}$ order sc corresponds to the 
approximation to the self-energy shown in Fig. 2d and given by 
Eq.(\ref{2scM}). The error bars on the Monte Carlo imaginary
time Green's function are not shown for clarity. They are of
the order of 0.001t.
(a) ${\cal G}({\bf k},\omega)$ at ${\bf k} = (\pi,0)$. 
(b) $A({\bf k},\omega)$ at ${\bf k} = (\pi,0)$. 
(c) ${\cal G}({\bf k},\omega)$ at ${\bf k} = (\pi/4,\pi/4)$. 
(d) $A({\bf k},\omega)$ at ${\bf k} = (\pi/4,\pi/4)$. 
(e) $N(\omega)$.}}
\end{figure}
\clearpage

\begin{figure}[ht]
\protect{\includegraphics*[width=\columnwidth]{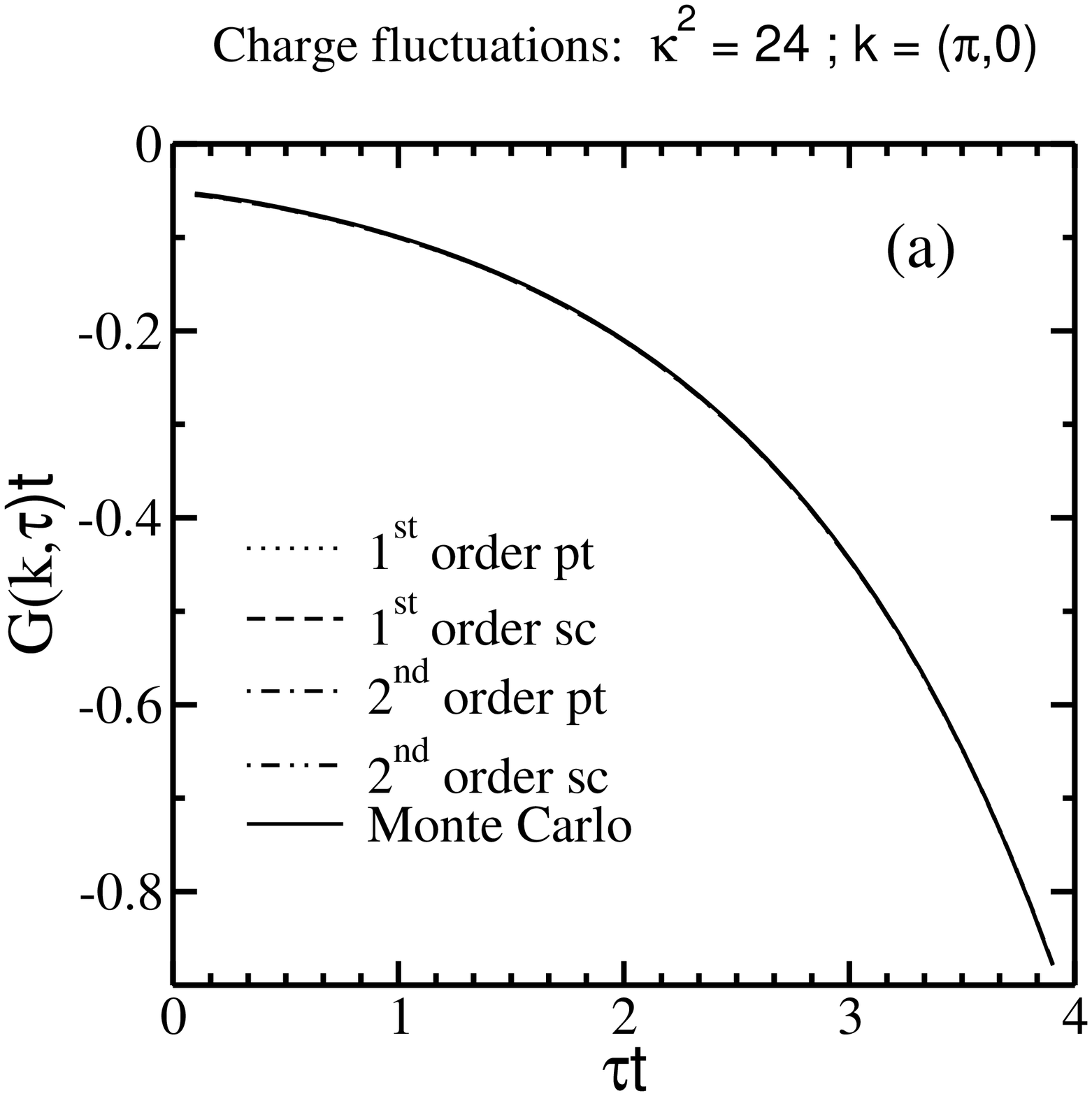}}
\end{figure}
\begin{figure}[ht]
\protect{\includegraphics*[width=\columnwidth]{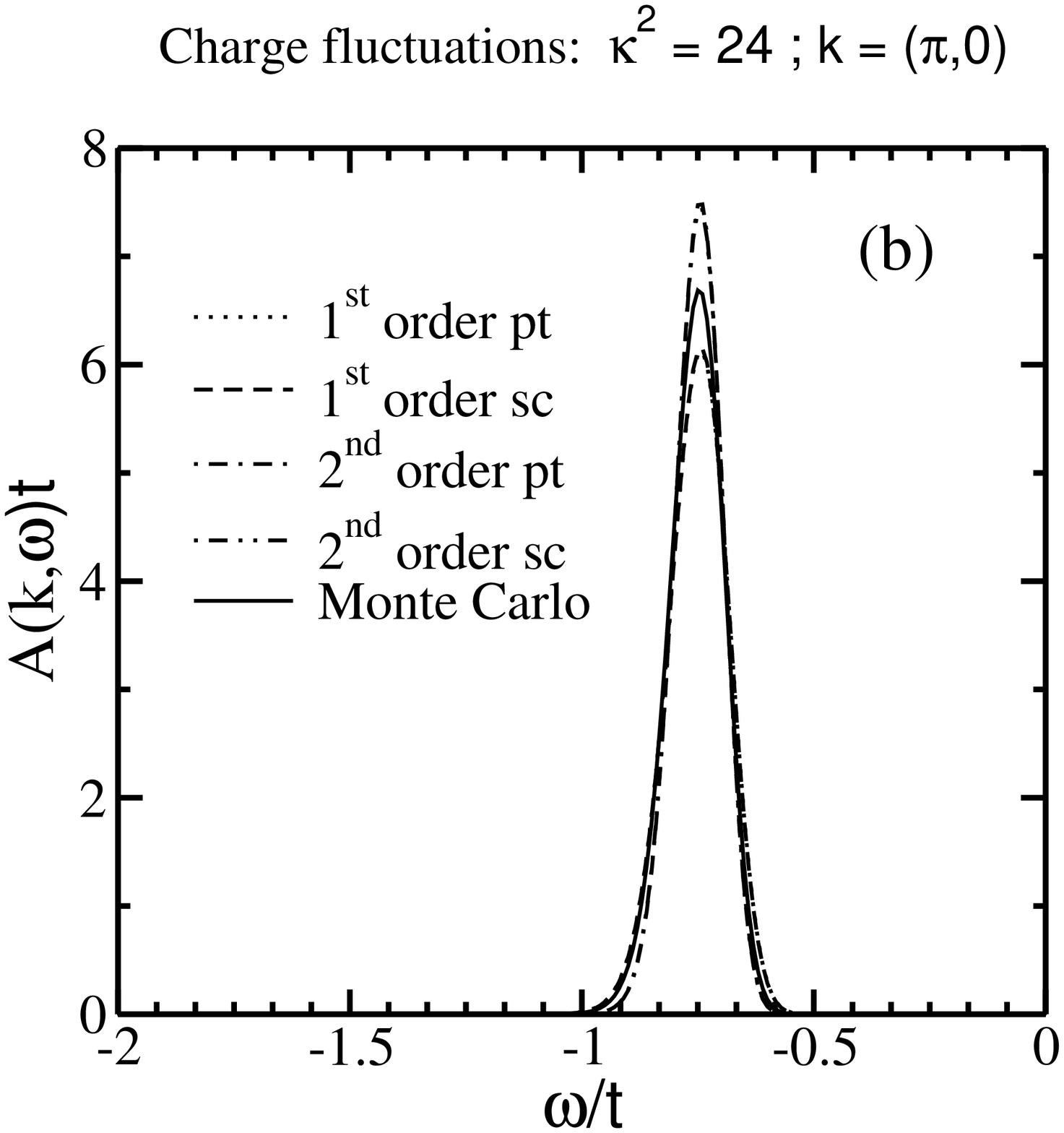}}
\end{figure}
\begin{figure}[ht]
\protect{\includegraphics*[width=\columnwidth]{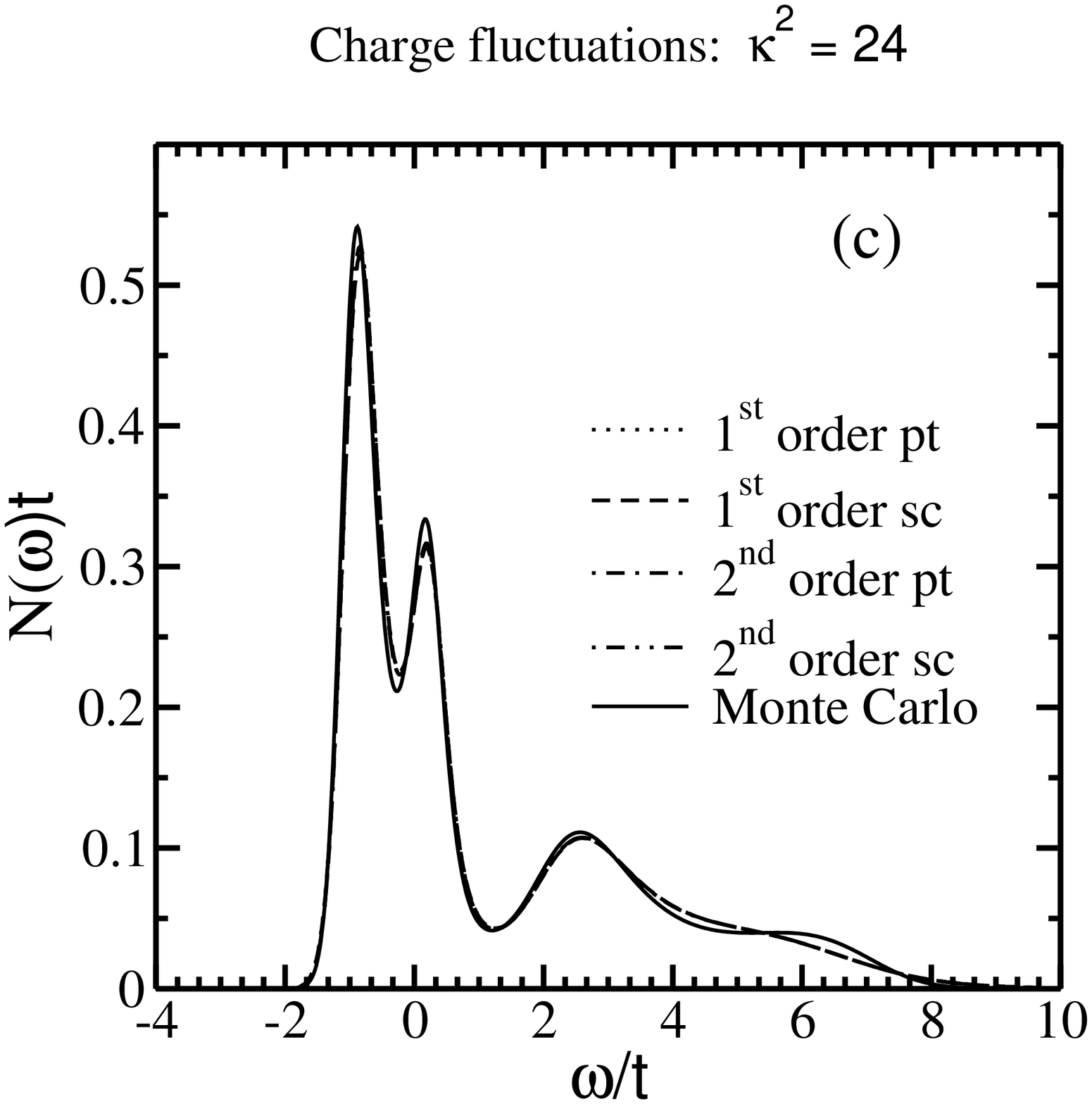}}
\protect{\caption{The various perturbation-theoretic approximations to
the quasiparticle imaginary time Green's function ${\cal G}({\bf k},\tau)$,
the spectral function $A({\bf k},\omega)$ and tunneling density
of states $N(\omega)$ are compared to the results of the Monte
Carlo simulations for $\kappa^2 = 24$ in the case of quasiparticles
coupled to charge fluctuations. $1^{st}$ order pt
corresponds to the approximation to the self-energy shown in Fig. 2a
and given by Eq.(\ref{1pt}). $1^{st}$ order sc corresponds to the 
approximation to the self-energy shown in Fig. 2b and given 
by Eq.(\ref{1sc}). The error bars on the Monte Carlo imaginary
time Green's function are not shown for clarity. They are of
the order of 0.00002t.
(a) ${\cal G}({\bf k},\omega)$ at ${\bf k} = (\pi,0)$. 
(b) $A({\bf k},\omega)$ at ${\bf k} = (\pi,0)$. 
(c) $N(\omega)$.}}
\end{figure}
\clearpage

\begin{figure}[ht]
\protect{\includegraphics*[width=\columnwidth]{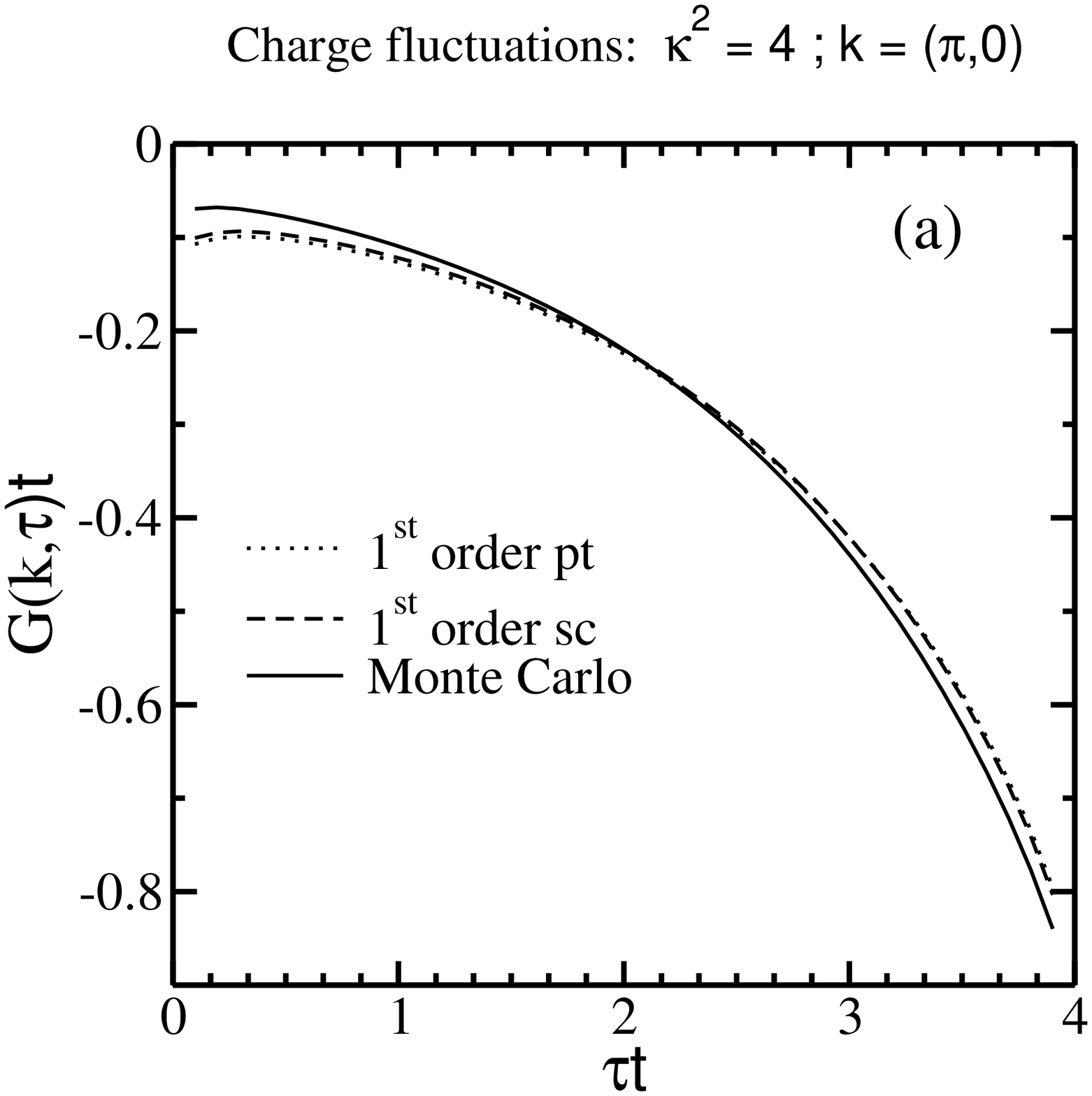}}
\end{figure}
\begin{figure}[ht]
\protect{\includegraphics*[width=\columnwidth]{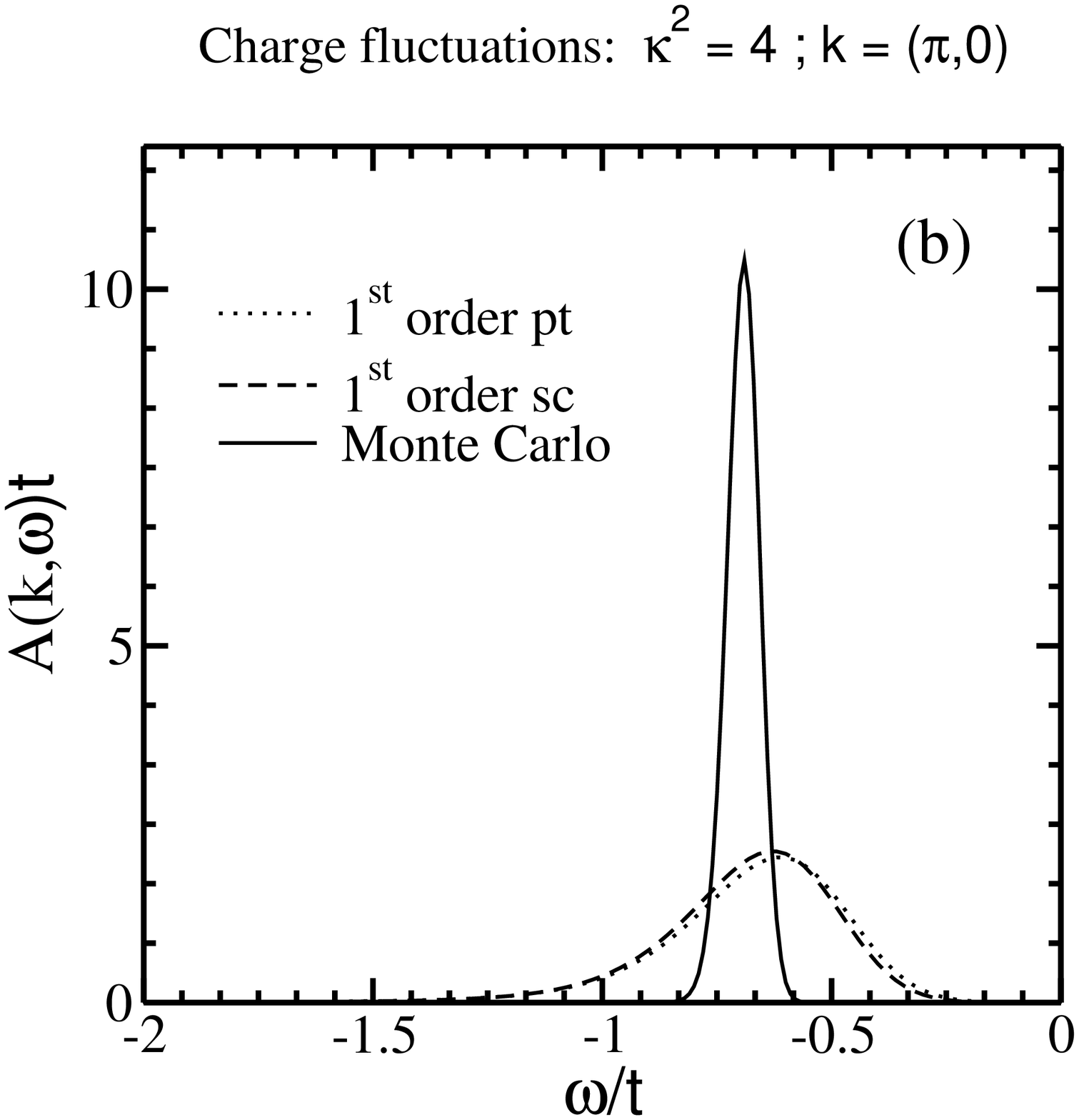}}
\end{figure}
\begin{figure}[ht]
\protect{\includegraphics*[width=\columnwidth]{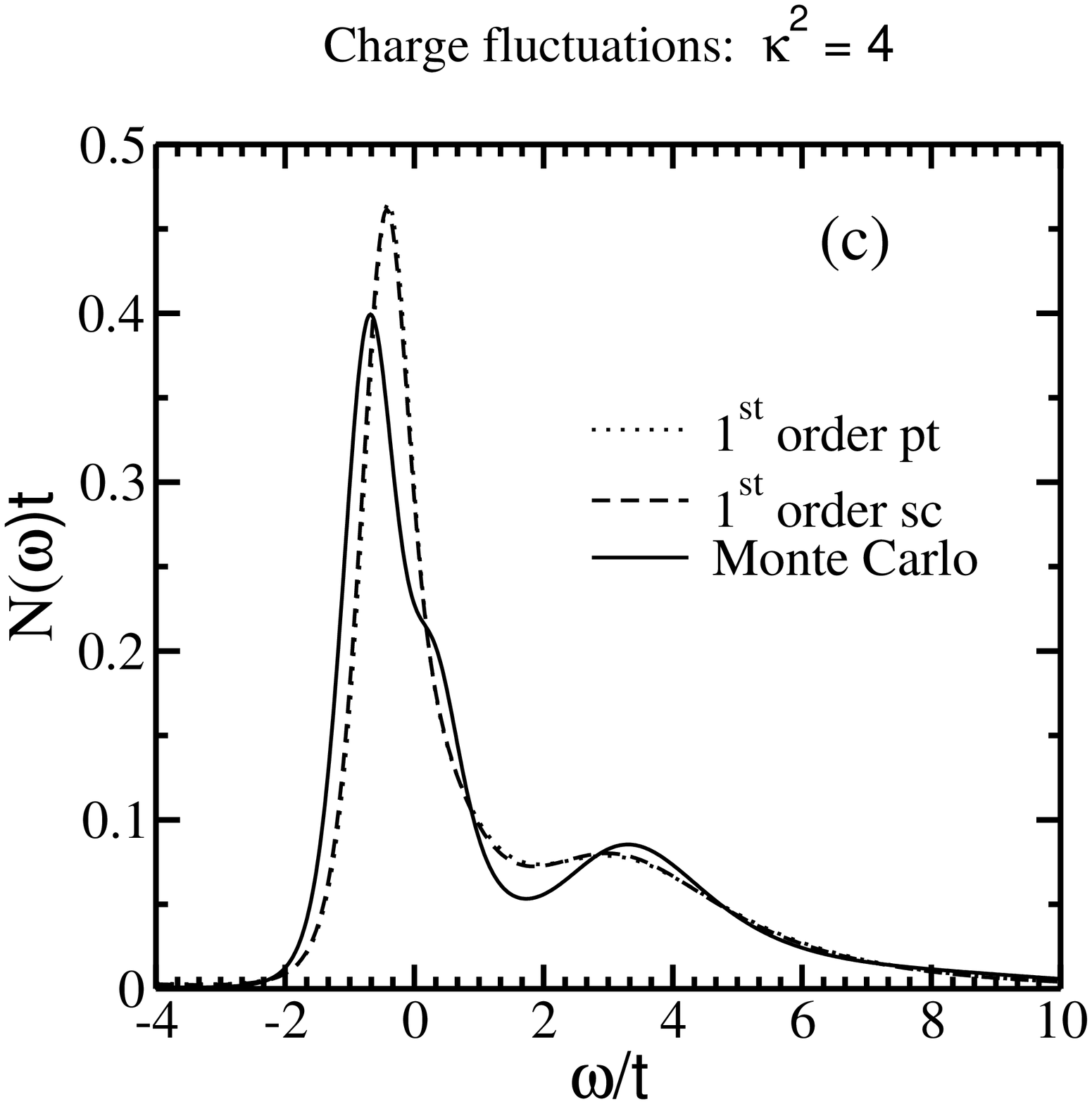}}
\protect{\caption{The various perturbation-theoretic approximations to
the quasiparticle imaginary time Green's function ${\cal G}({\bf k},\tau)$,
the spectral function $A({\bf k},\omega)$ and tunneling density
of states $N(\omega)$ are compared to the results of the Monte
Carlo simulations for $\kappa^2 = 4$ in the case of quasiparticles
coupled to charge fluctuations. $1^{st}$ order pt
corresponds to the approximation to the self-energy shown in Fig. 2a
and given by Eq.(\ref{1pt}). $1^{st}$ order sc corresponds to the 
approximation to the self-energy shown in Fig. 2b and given 
by Eq.(\ref{1sc}). The error bars on the Monte Carlo imaginary
time Green's function are not shown for clarity. They are of
the order of 0.0002t.
(a) ${\cal G}({\bf k},\omega)$ at ${\bf k} = (\pi,0)$. 
(b) $A({\bf k},\omega)$ at ${\bf k} = (\pi,0)$. 
(c) $N(\omega)$.}}
\end{figure}
\clearpage

\begin{figure}[ht]
\protect{\includegraphics*[width=\columnwidth]{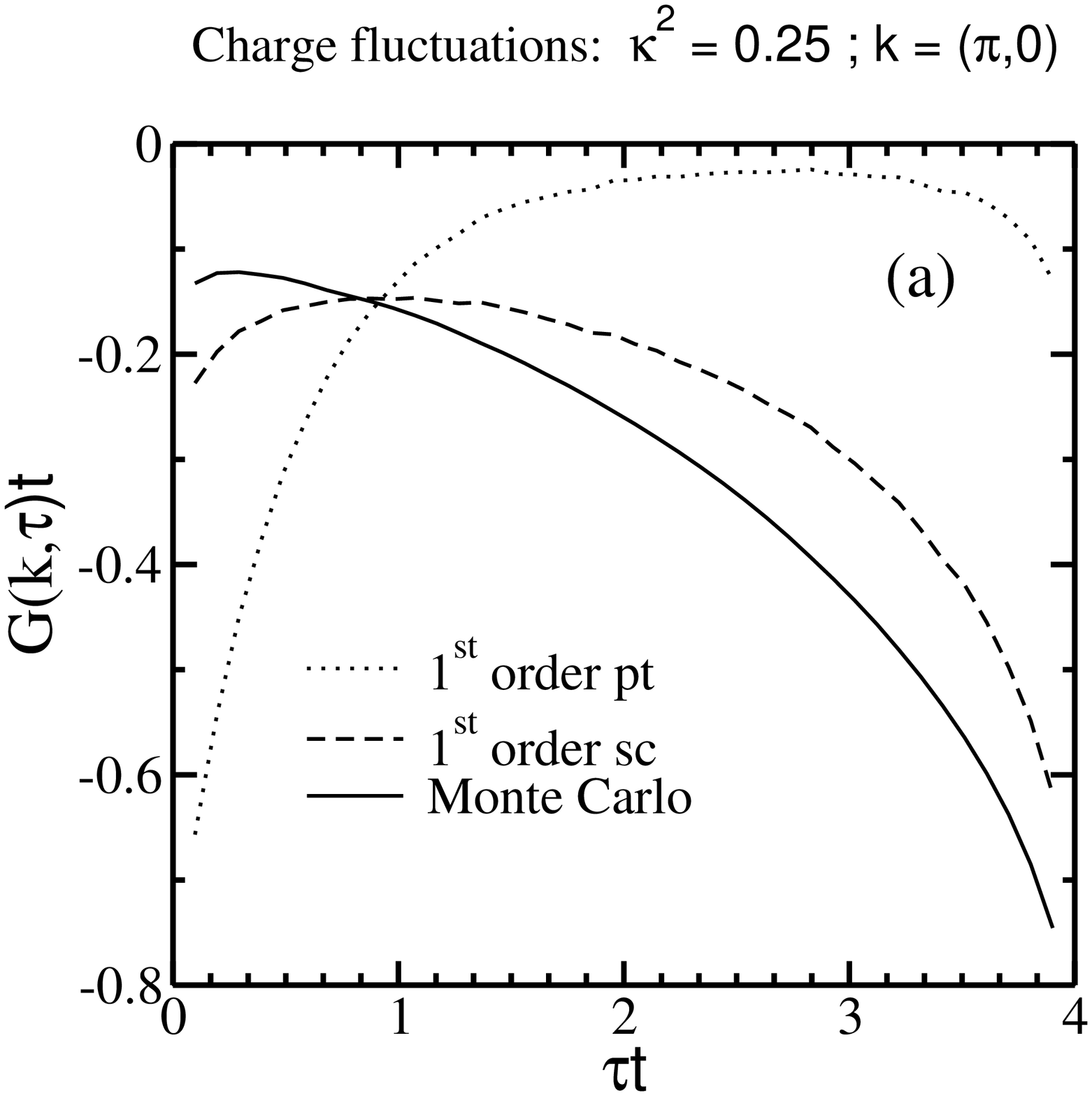}}
\end{figure}
\begin{figure}[ht]
\protect{\includegraphics*[width=\columnwidth]{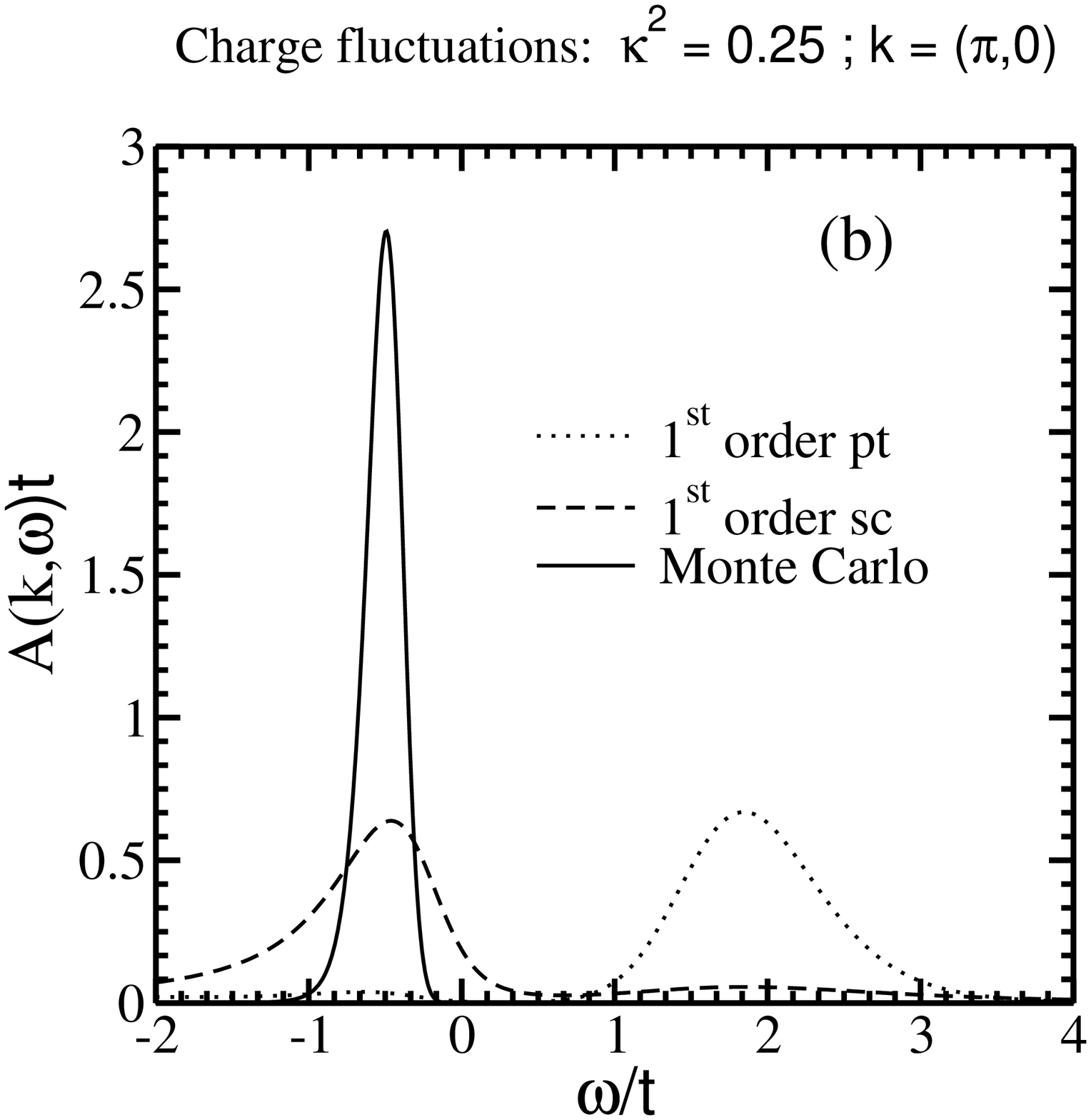}}
\end{figure}
\begin{figure}[ht]
\protect{\includegraphics*[width=\columnwidth]{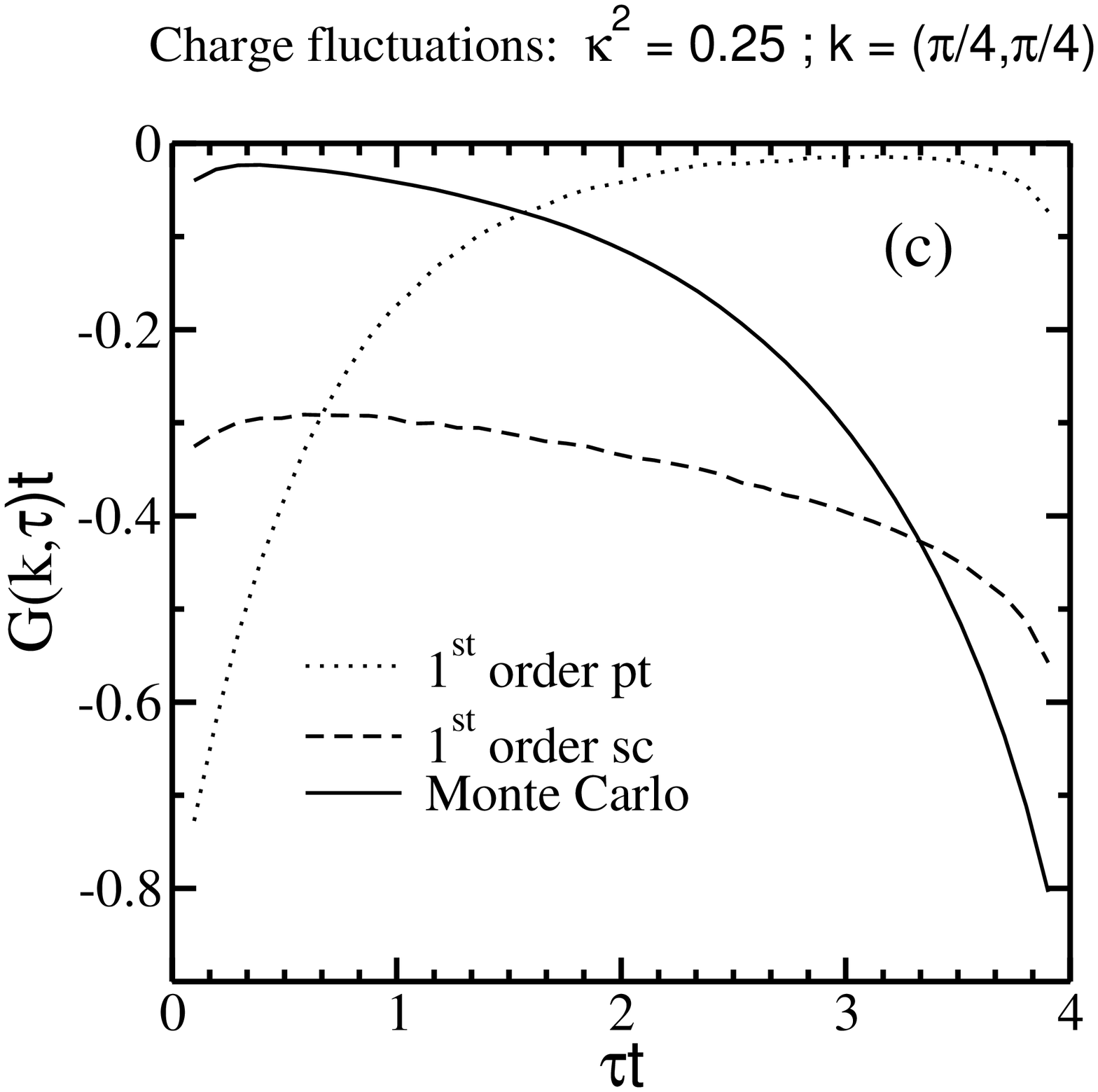}}
\end{figure}
\begin{figure}[ht]
\protect{\includegraphics*[width=\columnwidth]{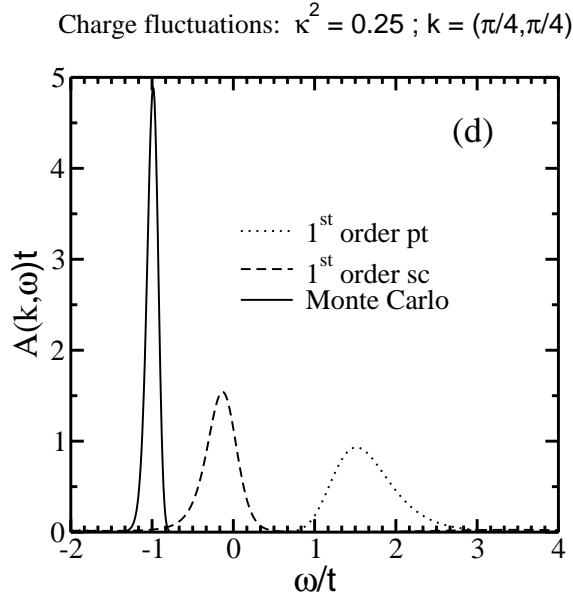}}
\protect{\caption{The various perturbation-theoretic approximations to
the quasiparticle imaginary time Green's function ${\cal G}({\bf k},\tau)$,
the spectral function $A({\bf k},\omega)$ are compared 
to the results of the Monte Carlo simulations for 
$\kappa^2 = 0.25$ in the case of quasiparticles
coupled to charge fluctuations. $1^{st}$ order pt
corresponds to the approximation to the self-energy shown in Fig. 2a
and given by Eq.(\ref{1pt}). $1^{st}$ order sc corresponds to the 
approximation to the self-energy shown in Fig. 2b and given 
by Eq.(\ref{1sc}). The error bars on the Monte Carlo imaginary
time Green's function are not shown for clarity. They are of
the order of 0.002t.
(a) ${\cal G}({\bf k},\omega)$ at ${\bf k} = (\pi,0)$. 
(b) $A({\bf k},\omega)$ at ${\bf k} = (\pi,0)$. 
(c) ${\cal G}({\bf k},\omega)$ at ${\bf k} = (\pi/4,\pi/4)$. 
(d) $A({\bf k},\omega)$ at ${\bf k} = (\pi/4,\pi/4)$.} }
\end{figure}
\clearpage

\end{document}